\documentclass[Afour,sageh,times]{sagej}

\usepackage{moreverb,url}

\usepackage[colorlinks,bookmarksopen,bookmarksnumbered,citecolor=red,urlcolor=red]{hyperref}
\usepackage{graphicx}
\usepackage{subfig}

\newcommand\BibTeX{{\rmfamily B\kern-.05em \textsc{i\kern-.025em b}\kern-.08em
T\kern-.1667em\lower.7ex\hbox{E}\kern-.125emX}}

\def\volumeyear{2021}

\def\Bb   {{\bf{B}}}
\def\Db   {{\bf{D}}}
\def\Eb   {{\bf{E}}}
\def\Hb   {{\bf{H}}} 
\def\Heff {{\bf{H}_{\rm eff}}}
\def\Jb   {{\bf{J}}}
\def\Mb   {{\bf{M}}}

\def\half {{\frac{1}{2}}}

\usepackage{hyperref}

\begin{document}

\runninghead{Yao, Jambunathan, Zeng et al.}

\title{A Massively Parallel Time-Domain Coupled Electrodynamics-Micromagnetics Solver}

\author{Zhi Yao\affilnum{1}, Revathi Jambunathan\affilnum{1}, Yadong Zeng\affilnum{2} and Andy Nonaka\affilnum{1}}

\affiliation{\affilnum{1}Lawrence Berkeley National Laboratory, Berkeley, CA, USA\\
\affilnum{2}University of Minnesota Twin Cities, Minneapolis, MN, USA}

\corrauth{Zhi Yao, 
1 Cyclotron Road, Mailstop 50A3111, Berkeley, CA 94720}

\email{jackie$\_$zhiyao@lbl.gov}

\begin{abstract}
We present a new, high-performance coupled electrodynamics-micromagnetics solver for full physical modeling of signals in microelectronic circuitry.
The overall strategy couples a finite-difference time-domain (FDTD) approach for Maxwell's equations to a magnetization model described by the Landau-Lifshitz-Gilbert (LLG) equation.
The algorithm is implemented in the Exascale Computing Project software framework, AMReX, which provides effective scalability on manycore and GPU-based supercomputing architectures.
Furthermore, the code leverages ongoing developments of the Exascale Application Code, WarpX, primarily developed for plasma wakefield accelerator modeling.
Our novel temporal coupling scheme provides second-order accuracy in space and time by combining the integration steps for the magnetic field and magnetization into an iterative sub-step that includes a trapezoidal discretization for the magnetization.
The performance of the algorithm is demonstrated by the excellent scaling results on NERSC multicore and GPU systems, with a significant (59x) speedup on the GPU using a node-by-node comparison.
We demonstrate the utility of our code by performing simulations of an electromagnetic waveguide and a magnetically tunable filter.
%\MarginPar{after yesterdays discussion about the other paper, should we still say novel temporal coupling scheme  AJN: sure get rid of the word 'novel'}
\end{abstract}


\maketitle

\section{Introduction}
As microelectronic devices become increasingly integrated and miniaturized, new materials and technologies are being adopted into conventional circuit designs. 
Spintronics is a promising technology and an important candidate for upcoming computing architectures due to their capability of reducing energy consumption and providing non-volatile memory.
\cite{chappert2007} explained that in this new technology, instead of moving charges, spin orientation of magnetic materials, also known as magnetization, is used as the information carrier.
\cite{chappert2007} and \cite{garello2013} also reviewed current applications including magnetic sensors and magnetic random access memory (MRAM) devices for computer architectures.
Ultimately, ``spin currents'' could even replace charge currents for the transfer and processing of  information, allowing faster and lower-energy operations, which further motivates advancing the field of spin electronics. 
%The motivation for including coupling to micromagnetic physics comes from the emergence of magnetic devices in computing architectures.
However, to propel this nascent technology into mainstream microelectronics ecosystem, pre-fabrication analysis is critical to determine and optimize the performance of new designs.
%is burgeoning and requires more research before becoming mainstream, 
Modeling and simulations provide an accurate lens to guide the development and therefore are ubiquitous in industrial, academic, and laboratory settings. 
%But, existing software cannot accurately capture very different physical characteristics inherent in these devices involving spintronic technologies, as well as other areas of innovation, including 5G application, quantum circuitry which also leverage heterogeneous materials
But existing software cannot accurately capture the very different physical characteristics that span over disparate spatial and temporal scales. 
Therefore, a trial-and-error approach is often used to test the performance of spintronic devices, thereby increasing the cost and time for development.
In addition to spintronic technologies, other areas of innovation in next-generation microelectronics, including 5G application, quantum circuitry, and multi-ferroic systems, are also in pressing need for simulation tools that can adequately capture interactions of multiple physics and accurately predict device performance.
%are also in keen needs for simulation tools that can adequately capture interactions of multiple physics and accurately predict device performance.
%are can which inherently involve heterogeneous materials with very different physical characteristics also lack adequate simulation tools to capture crucial multiphysics behavior that affect device performance.
Thus, there is an active demand for improved fidelity of electromagnetic (EM) simulations via higher spatiotemporal resolution and/or enhanced numerical strategies to seamlessly couple different physical models used to describe relevant physics in various regions of the circuit.

Existing commercially-available simulation packages are generally not suitable to perform such modeling due to two major drawbacks.
First, the software and algorithms are often proprietary, offering users limited customization of algorithms or incorporation of new physics required to model these new devices. Additionally, existing black-box algorithmic implementations may not inform users of potential assumptions that may break the validity of the approach for realistic devices.
%Also, the algorithm is typically not accessible to users. 
Second, scalability is limited to smaller clusters, rather than multicore/GPU-based exascale systems.
One of the most widely-used EM software, Ansys HFSS (High Frequency Structure Simulator) package, offers a platform to solve versatile EM applications, with the functionality to read in user-defined magnetic permeability tensors. 
However, it does not allow users to tailor their own mathematical models to recover the dynamical nonlinear material behavior such as nonlinear magnetic properties. 
%define their own PDEs 
COMSOL Multiphysics software is another commercial package, which allows for customized mathematical models, however, it may not be easily scalable on GPUs, limiting its applicability. %\MarginPar{do we know this for sure or should we say to the best of our knowledge?}
Therefore, many researchers are actively developing their own software.

Most software model the spin behavior of micromagnetic materials in isolation. 
Object Oriented MicroMagnetic Framework (OOMMF), developed by \cite{OOMMF}, has been developed to simulate spin dynamics in complex structures.
%In the micromagnetics area, research oriented software, such as object oriented micromagnetic framework (OOMMF) \cite{OOMMF}, has been developed to simulate spin dynamics in complex structures.
Recently \cite{fu2015finite} have extended OOMMF with GPU implementation and has demonstrated 35x speed-up compared to CPUs. 
Other solvers that model micromagnetic material include mumax developed by \cite{mumax} and PC micromagnetic simulator (Simulmag) developed by \cite{oti1997simulmag}.
%The recent burgeoning of magnetic material-based computing and communication components has driven researchers to explore the behavior of magnetic spins under the effect of EM signals.
The recent burgeoning of magnetic material-based computing and communication components has necessitated a close examination of the behavior of magnetic spins that are strongly influenced by the EM signals from the circuit. 
This study requires a hybrid solver coupling the EM signals with magnetic spin behavior, i.e., Maxwell equations with the Landau-Lifshitz-Gilbert (LLG) equation describing the spin precession dynamics. 
Various approaches have been used to study this coupled physics. 
\cite{couture2017coupled} have integrated a finite-element micromagnetic model and an integral-equation EM model to simulate the effect of eddy currents on magnetic spins.
\cite{venugopal2019nonlinear} have built a micromagnetic algorithm which predicts nonlinear ferromagnetic resonance frequency (FMR) shift phenomenon under the effect of time-harmonic high-power microwave EM fields.
Note that the magnetostatic assumption ($\partial \Db / \partial t = 0 $ in Ampere's law) applied by both \cite{couture2017coupled} and \cite{venugopal2019nonlinear} limits such tools in describing wave properties in circuits such as impedance matching. 
Efforts have also been made to implement fully dynamical EM waves ($\partial \Db / \partial t \neq 0 $), which can accurately model realistic devices. 
\cite{yao2018multiscale} have developed an alternating-direction-implicit (ADI) based time-domain micromagnetic algorithm. 
%with implementation of fully dynamic EM wave propagation. 
Additionally, \cite{Vacus2005} have implemented a 2D explicit Maxwell-LLG coupled algorithm, followed by \cite{Aziz2009} proposing a full three-dimensional implementation.

In this paper, we present a new coupled fully-dynamic electrodynamics-micromagnetics solver that is fully open-source and portable from laptops to manycore/GPU exascale systems.
The core solver in our code is a finite-difference time-domain (FDTD) implementation for Maxwell's equations that has been adapted to conditions found in microelectronic circuitry.
Thus we have incorporated spatially-varying material properties, boundary conditions, and external sources to  model our target problems.
We have also developed a new numerical coupling scheme to incorporate the LLG model of magnetization to allow for regions in space containing magnetic materials.
The overall scheme is second-order accurate in space and time.
In order to achieve portability and performance on a range of platforms, our code leverages the developments of two DOE Exascale Computing Project (ECP) code frameworks.
First, the AMReX software library developed by \cite{zhang2019amrex,zhang2020amrex} is the product of the ECP co-design center for adaptive, structured grid calculations.
AMReX provides complete data structure and parallel communication support for massively parallel manycore/GPU implementations of structured grid simulations such as FDTD.
Second, the WarpX accelerator code developed by \cite{vay2018warp} is an ECP application code for modeling plasma wakefield accelerators and contains many features that have been leveraged by our code.
These features include core computational kernels for FDTD, an overall time stepping framework, and I/O.
Looking forward, we will take advantage of adaptive mesh refinement (AMR) support, embedded boundaries for advanced geometry, and machine learning workflows that are continually being updated and improved by the WarpX team.

The rest of this paper is organized as follows.
In Section \ref{sec:model} we describe the equations that govern the involved physics.
In Section \ref{sec:numerical}, the details for our numerical approach are elaborated.
In Section \ref{sec:software}, we provide more details on the software framework, scaling studies, and a comparison of host vs.~GPU performance.
In Section \ref{sec:validation} we describe three test problems that validate our approach and implementation. 
%The first is a numerical convergence test to demonstrate accurate coupling of the micromagnetics and Maxwell equations and our treatment of spatially inhomogeneous material properties.
The first is a numerical convergence test to demonstrate the spatial and temporal order of accuracy of our coupled EM-micromagnetic solver with spatially inhomogeneous material properties.
%\MarginPar{ The convergence test only validates the order of accuracy, no? The YIG case is the one that demonstrates accurate coupling. }
The second is an X-band air-filled waveguide which has known analytical solutions that are used to validate the EM propagation in confined structures.
The third is a magnetically tunable filter realized by the waveguide with a yttrium iron garnet (YIG) thin slab inserted longitudinally which validates the coupling between the EM module and the LLG micromagnetics module. 
Finally, in Section \ref{sec:conclusions} we summarize this work and discuss future directions.

\section{Mathematical Model}\label{sec:model}
We begin with the full-form of the dynamic Maxwell's equations, Ampere and Faraday's laws,
\begin{equation}
\nabla\times\Hb = \Jb + \frac{\partial\Db}{\partial t},\label{eq:Ampere}
\end{equation}
\begin{equation}
\nabla\times\Eb = -\frac{\partial\Bb}{\partial t}\label{eq:Faraday}
\end{equation}
where, $\Db = \epsilon\Eb$ is the electric displacement, $\Eb$ is the electric field, $\epsilon$ is the permittivity of the medium, $\Bb$ is the magnetic flux density, $\Hb$ is the magnetic field, and $\Jb=\sigma\Eb + \Jb_{\rm src}$ is
the electric current density with conductivity, $\sigma$, and external source, $\Jb_{\rm src}$.
The permittivity, $\epsilon$ is the product of vacuum permittivity, $\epsilon_0$, and unit-less relative permittivity, $\epsilon = \epsilon_0\epsilon_r$.
Similarly, the permeability, $\mu$, is the product of vacuum permeability, $\mu_0$ and unit-less relative permeability, $\mu = \mu_0\mu_r$.
In non-magnetic materials, the constitutive relation between $\Bb$ and $\Hb$ is simply $\Bb = \mu \Hb$.
In magnetic materials, permeability could be described using the constitutive relation $\Bb = \mu \Hb = \mu_0(\Hb + \Mb)$, where $\Mb$ is the magnetization (total magnetic moments per unit volume).
It is easily derived from the above relations that $\mu_r = 1 + \Mb / \Hb$.
Typically, the value of $\mu_r$ is close to 1 in non-magnetic materials due to their very low intrinsic magnetization. 
In contrast, as explained by \cite{lax1962microwave}, the value of $\mu_r$ in magnetic materials is usually much larger than 1, especially in paramagnetic and ferromagnetic materials, due to large magnetization $\Mb$ values.
Moreover, in magnetic materials, $\mu_r$ typically varies with frequency and the field intensity, i.e.~$\mu_r$ has a dispersive frequency spectrum and nonlinear dependency on $\Hb$. 
%\MarginPar{this should be a full stop no? end of sentence? or are two sentences linked in some way?}
%\MarginPar{Does Mb always oscillate? it could also be damped? should we just say Mb evolves as a function of Hb or H-effective.}
The reason is that the magnetization, $\Mb$, required to determine $\mu_r$ evolves and oscillates as a function of $\Hb$, which in turn is solved by the Maxwell's equations. 
Therefore, the value of the $\mu_r$ is not known a priori for magnetic materials and instead of simply specifying an approximate $\mu_r$ value as the user-defined material property, it is more precise to predict the macroscopic material permeability for magnetic materials. 
As a result, solving the evolution of magnetization $\Mb$ in magnetic materials is essential in our model to describe the dynamic material behavior under the effect of EM fields.

One of the most widely used equation of motion to describe the spinning electrons is the LLG equation, conceived by \cite{landau1992theory}. 
%\MarginPar{a citation to the LLG literature}
The LLG equation describing the evolution of magnetization, $\Mb$, is
\begin{equation}
\frac{\partial\Mb}{\partial t} = \mu_0 \gamma (\Mb \times \Heff) + \frac{\alpha}{|\Mb|} \Mb \times \frac{\partial\Mb}{\partial t} + \boldsymbol{\tau},
\label{eq:LLG}
\end{equation}
where $\gamma$ represents the gyromagnetic ratio, with the value of $\gamma = -1.759 \times 10^{11}$ C/kg (note the negative sign indicating the opposite directions between the magnetic moment and the electron spin) for electrons. 
%\MarginPar{is this the gamma value for electrons for all LLG applications? }
The term $\alpha$ is the Gilbert damping constant, with typical values on the order of $10^{-5}$ to $10^{-3}$ (note the positive sign indicating the decreasing precession angle of $\Mb$). 
%\MarginPar{is there a citation for these values?}
It is also worth noting that the Gilbert damping constant, $\alpha$, encompasses all the physical mechanisms that lead to a damped $\Mb$, namely, lattice vibrations, spin-spin relaxation, spin-orbit coupling, and others. 
We chose to use the LLG form as the governing equation for magnetization evolution because it describes the overall damping process adequately and can be introduced in the magnetic susceptibility expression by a simple mathematical procedure, summarized by \cite{lax1962microwave}. 
$\Heff$ in \eqref{eq:LLG} is the effective magnetic field that drives the magnetic spins and can be decomposed into several terms accounting for the contributions from the oscillating EM field $\Hb$ (denoted as $\Hb_\mathrm{EM}$ from now on) as well as external static bias $\Hb_0$, anisotropy $\Hb_\mathrm{ani}$, and exchange coupling $\Hb_\mathrm{exch}$. 
%\MarginPar{what is the difference between $\Hb$ and $\Hb_\mathrm{EM}$? }
As the first demonstration of the model capability, we considered only the external DC bias $\Hb_0$ and the oscillating EM field $\Hb_\mathrm{EM}$.
At this point, it is straightforward to categorize the first two terms in the LLG equation:
the first term, $\mu_0 \gamma (\Mb \times \Heff)$, is the torque term which generates an angular momentum vector tangential to the spinning orbit of $\Mb$;
the second term, $(\alpha/|\Mb|) \Mb \times (\partial\Mb / \partial t) $, introduces a damping vector perpendicular to $\Mb$ and points toward the central axis of the $\Mb$ precession trajectory. 
Note that neither the torque term nor the Gilbert damping term influences the magnitude of the magnetization vector, $| \Mb |$. 
%\MarginPar{say 'neither the torque term nor the Gilbert damping term influces the magnitude of M' ?}
%For conciseness, only external bias and EM oscillation are included for the code capacity demonstration. 
The additional ``torque'' term, $\boldsymbol{\tau}$, expresses physics that is unique to spintronic devices, and takes various forms depending on the scenarios, such as the spin-transfer-torque(STT)-based structures. 
In this work, which is aimed to provide demonstration of the code essential capability, we focus only on the framework for coupling magnetization with EM signals and omit the $\boldsymbol{\tau}$ term to reduce complexity of our validation tests.
In future works, we will implement these terms to include more physics, specifically, $\Hb_\mathrm{ani}$, $\Hb_\mathrm{exch}$, $\boldsymbol{\tau}$, and so on, as would be needed to recover physics such as magnetic domain movement, spin waves, STT, and others.
%\MarginPar{here we say that the tau terms dominate spatially varying magnetization.. we also have spatially varying magnetization without modeling torque. sorry for being confused about this. }
In fact, the more complex spatial variation of the magnetization in the ferromagnets comes mainly from the these terms.
Under the assumption of including only EM coupling and DC bias as mentioned above, the magnitude of $\Mb$ stays constant.
Since in both RF and memory applications, magnetization is mostly biased to saturation, it is reasonable to use $|\Mb|=M_s$, where $M_s$ is a spatially-dependent material property representing the saturation level.
%\MarginPar{so -- stupid question : Ms is constant in time, but can vary in space? and that is how we can recognize which regions are magnetic and non-magnetic?}

%\MarginPar{I moved these alternate Faraday equations from the numerical section to this model section since then we are referring to the actual equations that we are solving in this paragraph}
In our numerical approach, we choose $\Hb$ as the primary connection between the EM module and micromagnetics module, as $\Hb$ affects both electric currents and magnetization as discussed previously.
Thus, we use a modified version of Faraday's law that has $\Bb = \mu_0(\Hb + \Mb)$ substituted in ~\eqref{eq:Faraday} in magnetic regions to give
\begin{equation}
\frac{1}{\mu_0}\nabla\times\Eb = -\frac{\partial\Hb}{\partial t} - \frac{\partial\Mb}{\partial t}.\label{eq:Faraday2}
\end{equation}
and in non-magnetic regions,
\begin{equation}
\frac{1}{\mu}\nabla\times\Eb = -\frac{\partial\Hb}{\partial t}\label{eq:Faraday3}
\end{equation}

To summarize, we have built the coupled model to include spatial inhomogeneity of materials, containing both magnetic and non-magnetic regions.
In magnetic regions, our coupled algorithm solves both Maxwell's equations (\ref{eq:Ampere}), (\ref{eq:Faraday2}) and the LLG equation (\ref{eq:LLG}). 
In non-magnetic regions, the simple constitutive relation $\Bb = \mu \Hb$ is applied and only Maxwell's equations (\ref{eq:Ampere}) and (\ref{eq:Faraday3}) are solved. 
This is an important distinction as our algorithm is designed to model devices containing both magnetic and non-magnetic regions.
%We also note that we can analyze the simulation results to \textcolor{red}{obtain} an effective $\mu$ \textcolor{red}{for the magnetic materials} by converting the field variables into the frequency domain and using the definition $\mu_r = 1 + \Mb / \Hb = \Bb / \mu_0 \Hb $. 
%\MarginPar{we can do this only in the magnetic regions, no? Do we actually compute this for the tunable filter? if not, since we mention this sentance here, it might induce curiosity to see what the mu from our sim? }

\section{Numerical Approach}\label{sec:numerical}

\subsection{Spatial Discretization}
\begin{figure}[tb]
\centering
\includegraphics[width=0.45\textwidth]{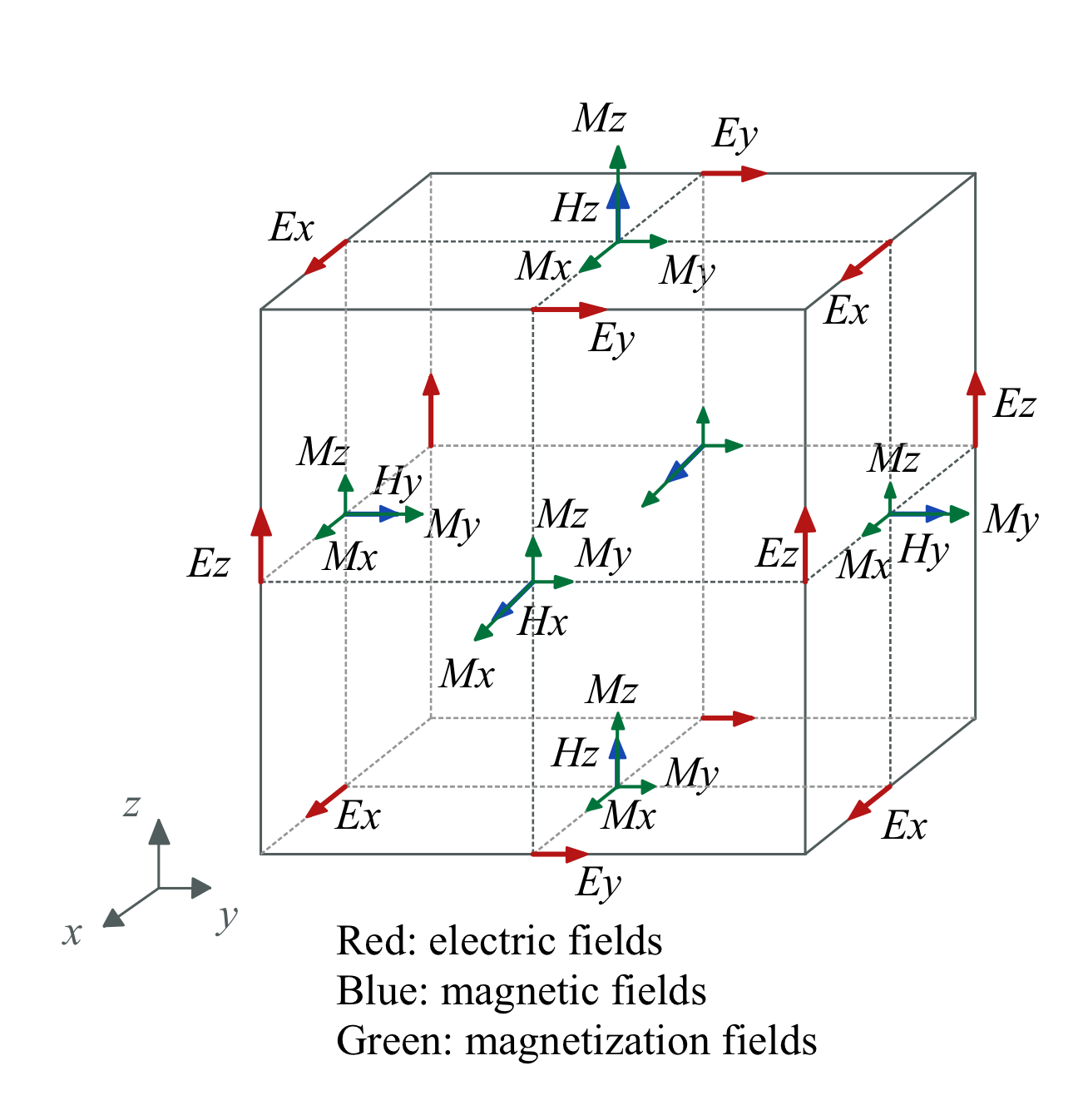}
\caption{Yee's grid with normal magnetic field on cell faces, tangential electric field on cell edges, and all three components of the magnetization at each cell faces.}
\label{fig:yee}
\end{figure}

%\sout{We use a uniform Cartesian mesh, where each cuboidal (not necessarily cubic) grid cell has side lengths $\Delta x, \Delta y$, and $\Delta z$.}
We use a uniform Cartesian mesh with cell-sizes $\Delta x, \Delta y$, and $\Delta z$.
%\sout{A given cell has a cell-center, 6 faces (nodal in one direction), 12 edges (nodal in two directions), and 8 nodes (nodal in all three directions).}
We employ the standard \cite{yee1966numerical} grid configuration for electrodynamics where the normal components of $\Hb$ fields are defined on cell-faces, and the tangential components of $\Eb$ and $\Jb$ are defined on the cell-edges as illustrated in Fig.~\ref{fig:yee}.
%Specifically, the normal components of $\Hb$ are placed on normal faces, and the transverse components of $\Eb$ and $\Jb$ are placed on edges (e.g., $E_x$ is placed on an edge that is nodal in the $y$ and $z$ directions)\textcolor{red}{, as illustrated in Fig.~\ref{fig:yee}.}
The magnetization, $\Mb$, is also defined on cell-faces, but with an important distinction.
At each face, all three components of $\Mb$ are stored, as opposed to $\Hb$ and $\Bb$, where only the normal component is stored.
%Refer to Fig. \ref{fig:yee}.

Using this approach, the curl of $\Hb$ can be simply and naturally represented with second-order accuracy at the locations of $\Eb$, and vice versa.
At various points in the algorithm, transverse components of $\Hb$ are required on faces (e.g., $H_y$ on an $x$-face).
In these cases, we compute these values using a second-order four-point average of the nearest values.
The material properties ($\epsilon, \mu, M_s, \sigma, \gamma, \alpha$) are specified through user-defined functions of space and/or time supplied at run-time using the function parser capability of the code.
To compute the material properties, we first evaluate them from the parsed function point-wise at cell centers and then later interpolate these values to other locations on the grid as required using second-order stencils.
By making this choice we are using the interpretation that any given computational cell can be thought of as a finite volume cell consisting of a material with given properties; 
in the future as we consider more complex, non-grid-aligned geometries, we will consider the viability of using a volume fraction weighted parameterization of the material, and compare that approach to more advanced discretizations involving embedded boundaries.
The algorithm as presented here requires $\sigma$ and $\epsilon$ on edges where $\Eb$ and $\Jb$ are located, and $\gamma, \alpha$, and $M_s$ on faces where $\Mb$ and $\Hb$ are located.
Note that we also parse the initial condition for $\Mb$ at cell centers and average these values to faces; in this way there is a consistent definition of the material property $M_s$ and the initial field $\Mb$ that allows the initial condition $|\Mb|=M_s$ to hold discretely.
The $\Eb$ and $\Hb$ fields are initialized with user-defined functions directly onto the locations of these fields on the yee-grid, as are any external sources corresponding to these fields.

\subsection{Temporal Discretization}
%\MarginPar{Need to discuss here (or in intro) what other people, including Jackie, have previously done to solve these equations (i.e., ADI) and why we are doing what we are doing}
%\MarginPar{add note on why we aren't implementing Jackie's ADI.  Dispersion.  Expensive linear solvers.  Maybe for the future}
Explicit time-domain algorithm is implemented in this work as it leads to less numerical dispersion errors compared to implicit algorithms proposed by \cite{Zheng2000,Namiki1999,yao2018multiscale}, albeit its computational cost. 
% \sout{Although being more computationally expensive, explicit FDTD leads to less numerical dispersion errors compared to implicit time-domain algorithms.} 
For future developments, we may explore the option of implementing implicit algorithms to improve computational efficiency by employing larger time-steps compared to explicit methods, with the recognition of challenges in the GPU implementation and field boundary issues described by \cite{yao2018multiscale}.
%\MarginPar{is over-sampling a widely used term in microelectronics? coming from plasma, I understand sampling in terms of particles. is over-sampling a bad thing? } 
The overall scheme resembles the well-known second-order leapfrog scheme used to advance $\Hb$ and $\Eb$ in standard FDTD schemes.
The temporal discretization adopted is similar to that described by \cite{Aziz2009}.
In this scheme, at every time step, $\Hb$ is first advanced over a half time step ($\Hb^{n}\rightarrow\Hb^{n+1/2}$), followed by a full time step update of $\Eb$ ($\Eb^{n}\rightarrow\Eb^{n+1}$), concluding with an additional half time step update of $\Hb$ ($\Hb^{n+1/2}\rightarrow\Hb^{n+1}$).
We again note that in magnetic regions, we choose to update both $\Hb$ and $\Mb$ simultaneously using ~(\ref{eq:Faraday2}) and a second-order accurate iteration scheme which satisfies the LLG equation (\ref{eq:LLG}) to a user-specified tolerance.
%\textcolor{red}{Note that the code updates $\Bb$ through the constitutive relation, but it does not couple to any of the other variables anymore.  Do we want to comment on this and/or analyze to what extent this $\Bb$ compares to $\Bb$ computed with $\Bb=\mu_0(\Hb+\Mb)$?}

As mentioned in Section \ref{sec:model}, analytically, the magnitude of $\Mb$ at each point of space cannot change since the terms in the right-hand-side are orthogonal to the direction of $\Mb$.
However, since we are using finite-sized time steps, numerically, the magnitude of $\Mb$ will slightly change with each time step.
Thus, after each update of $\Mb$, we renormalize $\Mb$ using the function $\mathcal{N}(\Mb) \equiv \Mb \times (M_s / |\Mb|)$.
The renormalization step is included in the algorithm descriptions below.

Additionally, spatially-varying, time-dependent soft-source and hard-source external excitations are supported for both the $\Eb$ and $\Hb$ fields.
\cite{taflove2005computational} defines soft sources in terms of the time-rate of field change, and are applied at the beginning of the time step.
Hard sources are enforced directly for the corresponding fields after their updates at each time step.

\subsubsection{First-Order Temporal Scheme}\label{sec:first-order}
Although the focus of this paper is the second-order scheme, and the majority of simulations are not performed using this first-order scheme, describing this method in full allows for an easier understanding of how we implemented the second-order method.
In Section \ref{sec:second-order} we will discuss the modifications to the first-order scheme that are required for second-order temporal accuracy.
%\sout{These modifications consist of iterative schemes that replace {\bf Steps 1} and {\bf 3} to \sout{integrate} \textcolor{red}{discretize} $\Hb$ and $\Mb$ to second-order accuracy.}
An analytically equivalent form of the LLG equations described in ~(\ref{eq:LLG}) is, 
%\MarginPar{a citation to refer to this would be great here. JY --- citation added after the equation}
\begin{equation}
\frac{\partial\Mb}{\partial t} = \mu_0 \gamma_{\rm{L}} (\Mb \times \Heff) + \frac{\alpha \mu_0 \gamma_{\rm{L}}} {| \Mb |} \Mb \times (\Mb \times \Heff) ,
\label{eq:LLG_scalar}
\end{equation}
where $\gamma_L = \gamma / (1 + \alpha ^2)$ \cite{lax1962microwave}, and $\boldsymbol{\tau}$ has been omitted. 
This formulation will be used in our first-order temporal scheme. \\ \\
%\MarginPar{we use this formulation for both first and second order, no?}\\ \\
{\bf Step 1:} We compute $\Mb^{n+\half}$ by reformulating the scalar form of the LLG equation (\ref{eq:LLG_scalar}) using a forward-Euler discretization proposed by \cite{Vacus2005},
%\begin{equation}
\begin{multline}
\Mb^{n+\half} = \Mb^n \\
+ \Delta t\left[\mu_0 \gamma_{\rm{L}} (\Mb \times \Heff) + \frac{\alpha \mu_0 \gamma _{\rm{L}}} {M_s} \Mb \times (\Mb \times \Heff)\right]^n
\end{multline}
%\end{equation}
Note that the transverse components of $\Heff$ are spatially averaged to update the respective $\Mb$ components at the cell-faces. 
%\sout{Note that the integration of $\Mb$ is purely local in the sense that there are no spatial gradients or curls of any variables involved, although spatial averaging of $\Heff$ is still required to obtain the transverse components at cell-faces.} \MarginPar{I removed this because M cross H is a curl and could be misleading.}
We normalize $\Mb^{n+\half} = \mathcal{N}(\Mb^{n+\half})$ and then compute $\Hb^{n+\half}$ by solving Faraday's law (\ref{eq:Faraday2}) with a forward-Euler discretization of the electric field, 
%\MarginPar{just a thought : do we mention anywhere that this update is only in magnetic regions? We have mentioned it earlier in math model, but wondering if we should say compute Hb in magnetic regions by discretizing Faraday's law. JY: At the end of first paragraph of 3.2, we say M and H are only in magnetic regions; and at the very end of Section 2, we also mentioned this. At the end of this section ,we also mention Ms=0 region. So I think it should be pretty clear already?}
\begin{equation}
\frac{1}{\mu_0}\nabla\times\Eb^n = \frac{\Hb^n - \Hb^{n+\half}}{\Delta t / 2} + \frac{\Mb^n - \Mb^{n+\half}}{\Delta t / 2},
\end{equation}
which simplifies to
\begin{equation}
\Hb^{n+\half} = \Hb^n + \Mb^n - \Mb^{n+\half} - \frac{\Delta t}{2\mu_0}\nabla\times\Eb^n.\label{eq:H_update}
\end{equation}
In nonmagnetic regions, i.e.~in regions where $\Mb=0$, considering Faraday's law (\ref{eq:Faraday3}) is sufficient to compute $\Hb$. Therefore, in regions where $M_s=0$, we simply set $\Mb=0$ and integrate $\Hb$ using (\ref{eq:H_update}) but with $\mu_0$ replaced by the local value of $\mu$, which reduces to the form described in (\ref{eq:Faraday3}).\\ \\
{\bf Step 2:} We compute $\Eb^{n+1}$ by discretizing Ampere's law (\ref{eq:Ampere}) using time-centered representations of the magnetic field and current,
%\begin{equation}
\begin{multline}
\nabla\times\Hb^{n+\half} = \half\left(\sigma\Eb^n + \sigma\Eb^{n+1}\right) \\
+ \Jb_{\rm src}^{n+\half} + \epsilon\frac{\Eb^{n+1} - \Eb^n}{\Delta t},
\end{multline}
%\end{equation}
which simplifies to 
%\MarginPar{we possibly need a sentance somewhere that says that our material properties are spatially varying scalars}
%\begin{equation}
\begin{multline}
\Eb^{n+1} = \left(\frac{\sigma}{2} + \frac{\epsilon}{\Delta t}\right)^{-1} \\
\times \left[\nabla\times\Hb^{n+\half} - \left(\frac{\sigma}{2} - \frac{\epsilon}{\Delta t}\right)\Eb^n - \Jb_{\rm src}^{n+\half}\right].
\end{multline}
%\end{equation}
%\textcolor{red}{where both $\sigma$ and $\epsilon$ are spatially varying scalars... AJN: redundant since we mentioned this in spatial discretization.  And if we note that these are spatially varying here, we should do the same for all material properties in the H/M equations, which adds even more redundancy}
{\bf Step 3:} We compute $\Mb^{n+1}$ following a similar procedure described in {\bf Step 1}, using a forward-Euler discretization of the LLG equation (\ref{eq:LLG_scalar}):
%\begin{equation}
\begin{multline}
\Mb^{n+1} = \Mb^{n+\half} \\
+ \Delta t\left[\mu_0 \gamma_{\rm{L}} (\Mb \times \Heff) + \frac{\alpha \mu_0 \gamma _{\rm{L}}} {M_s} \Mb \times (\Mb \times \Heff)\right]^{n+\half}
\end{multline}
%\end{equation}
We normalize $\Mb^{n+1} = \mathcal{N}(\Mb^{n+1})$ and then compute $\Hb^{n+1}$ by integrating Faraday's law (\ref{eq:Faraday2}) with a backward-Euler discretization of the electric field,
\begin{equation}
\frac{1}{\mu_0}\nabla\times\Eb^{n+1} = \frac{\Hb^{n+\half} - \Hb^{n+1}}{\Delta t / 2} + \frac{\Mb^{n+\half} - \Mb^{n+1}}{\Delta t / 2},
\end{equation}
which simplifies to
\begin{equation}
\Hb^{n+1} = \Hb^{n+\half} + \Mb^{n+\half} - \Mb^{n+1} - \frac{\Delta t}{2\mu_0}\nabla\times\Eb^{n+1}.\label{eq:H_update2}
\end{equation}
Again, in nonmagnetic regions, we simply set $\Mb=0$ and integrate $\Hb$ using (\ref{eq:H_update2}) but with $\mu_0$ replaced by the local value of $\mu$.

\subsubsection{Second-Order Temporal Scheme}\label{sec:second-order}
We replace {\bf Step 1} and {\bf Step 3} described in the first-order method of Section \ref{sec:first-order} with an iterative scheme that utilizes a trapezoidal discretization of the vector form of the LLG equation (\ref{eq:LLG}).
In this algorithm, the exposition of {\bf Step 1} and {\bf Step 3} are identical, %\MarginPar{RJ : English question , what is the meaning of expositionally? JY: I know "expositional" but not sure if there is an adverb form.}
except that in {\bf Step 1} we advance $(\Hb,\Mb)$ from $t^n$ to $t^{n+\half}$ using the curl of $\Eb^n$, and in {\bf Step 3} we advance $(\Hb,\Mb)$ from $t^{n+\half}$ to $t^{n+1}$ using the curl of $\Eb^{n+1}$.
Thus, we only write out the details for {\bf Step 1}.
%\textcolor{red}{The second-order scheme follows the same spatial discretization as that described in Steps 1 and 3 of first-order scheme, however, to advance $(\Hb,\Mb)$ from $t^{n+\half}$ to $t^{n+1}$ we use the curl of $\Eb^{n+1}$ instead of $\Eb^{n}$ used in the above described first-order scheme.}
\\ \\
{\bf Step 1:}
We begin by setting $(\Hb,\Mb)^{n+\half,r=0} = (\Hb,\Mb)^n$, where $r$ is the counter for the iterative procedure. 
%Then we iterate the following over $r$ 
%\MarginPar{when we say iterate of $r$, I feel the reader would expect the range of $r$. so tried to re-word it.}
Then we iteratively solve (18)-(21) given below, beginning with $r=1$ until the change in $\Mb^{n+\half,r}$ from one iteration to the next is sufficiently small.
Specifically, the iterations stop when the maximum value of $(\Mb^{n+\half,r} - \Mb^{n+\half,r-1})/M_s$ over the entire domain is smaller than a user-defined tolerance that we choose to be $10^{-6}$.
Note $M_s$ is the local value at each corresponding mesh.
%\textcolor{red}{To determine convergence for the method in the solver, we chose a value of $\delta=XYZ$}.
Consider a second-order, trapezoidal discretization of (\ref{eq:LLG}),
%\begin{equation}
\begin{multline}
\frac{\Mb^{n+\half} - \Mb^{n}}{\Delta t/2} = \\
\frac{\mu_0 \gamma}{2} (\Mb^{n} \times \Hb^{n}_{\rm eff}) 
+ \frac{\mu_0 \gamma}{2} (\Mb^{n+\half} \times \Hb^{n+\half}_{\rm eff}) \\ 
+ \frac{\alpha}{M_s}\left(\frac{\Mb^n + \Mb^{n+\half}}{2}\right) \times \frac{\Mb^{n+\half} - \Mb^{n}}{\Delta t/2}, \label{eq:llg_tra1}
\end{multline}
%\end{equation}
Since $\Mb^n \times \Mb^n = 0$, (\ref{eq:llg_tra1}) can be expressed as
\begin{multline} \label{eq:llg_tra2}
\frac{\Mb^{n+\half} - \Mb^{n}}{\Delta t/2} = \\
\frac{\mu_0 \gamma}{2} (\Mb^{n} \times \Hb^{n}_{\rm eff}) + 
\frac{\mu_0 \gamma}{2} (\Mb^{n+\half} \times \Hb^{n+\half}_{\rm eff}) \\
- \frac{\alpha}{M_s}\frac{(\Mb^{n+\half} \times \Mb^{n})} {\Delta t/2}.
\end{multline}
Reordering terms gives
\begin{multline} \label{eq:llg_tra3}
\Mb^{n+\half} = 
\Mb^{n} + \frac{\mu_0 \gamma \Delta t}{4} (\Mb^{n} \times \Hb^{n}_{\rm eff}) \\
-
\Mb^{n+\half}
\times
\left(-\frac{\mu_0 \gamma \Delta t}{4} \Hb^{n+\half}_{\rm eff} + 
\frac{\alpha} {M_s} \Mb^{n}\right).
\end{multline}
If we replace $\Mb^{n+\half}$ with $\Mb^{n+\half,r}$ and $\Hb^{n+\half}_{\rm eff}$ with $\Hb^{n+\half,r-1}_{\rm eff}$, we are essentially lagging some terms in the iterative update, noting the accuracy of the procedure improves with iteration count. Also, we can now directly solve for $\Mb^{n+\half,r}$ using the following analytical expression derived by \cite{Vacus2005},
\begin{equation} \label{eq:llg_tra4}
\Mb^{n+\half,r} = 
\frac{\boldsymbol{b} + (\boldsymbol{a} \cdot \boldsymbol{b}) \boldsymbol{a} - \boldsymbol{a} \times \boldsymbol{b}}{1+|\boldsymbol{a}|^{2}},
\end{equation}
where
\begin{equation} \label{eq:llg_tra5}
\boldsymbol{a} = - \left[ \frac{ \mu_0 |\gamma| \Delta t}{4} \Hb^{n+\half,r-1}_{\rm eff}
+ \frac{\alpha}{M_s}\Mb^n
\right],
\end{equation}
\begin{equation} \label{eq:llg_tra6}
\boldsymbol{b} = \Mb^{n} - \frac{\mu_0 |\gamma| \Delta t}{4} (\Mb^{n} \times \Hb^{n}_{\rm eff}). 
\end{equation}
We normalize $\Mb^{n+\half,r} = \mathcal{N}(\Mb^{n+\half,r})$ and then compute $\Hb^{n+\half,r}$ by integrating Faraday's law as in the first-order algorithm, with an updated approximation for the time-derivative of $\Mb$,
\begin{equation}
\Hb^{n+\half,r} = \Hb^n + \Mb^n - \Mb^{n+\half,r} - \frac{\Delta t}{2\mu_0}\nabla\times\Eb^n.\label{eq:iterate_H}
\end{equation}
We continue to iterate over $r$ using (\ref{eq:llg_tra4}) and (\ref{eq:iterate_H}).
%\MarginPar{Another question -- we say earlier that we stop iteration when $H^r$ and $M^r$ change negligibly. But we actually test convergence only by checking the change in $M$. Its a minor point, but may be good to be consistent in the write-up}
After the iterations converge, we set $(\Hb,\Mb)^{n+\half} = (\Hb,\Mb)^{n+\half,r}$ and proceed to {\bf Step 2}. 
Note that in regions where $M_s=0$, we set $\Mb=0$, skip this iterative scheme, and instead integrate $\Hb$ using (\ref{eq:iterate_H}) but with $\mu$ replacing $\mu_0$.
%Again, this is a discretization of equation (\ref{eq:Faraday3}).

\subsubsection{Boundary Conditions}
%\MarginPar{Jackie, Andy : does this read okay? Do we need more details? AJN: This sounds good; do you want to mention PMC even though we don't use it here? Reva: it is WIP and currently being added. can I still mention it? if so, I will add it in the context of a symmetry boundary?}
Our solver supports three types of boundary conditions that are required to ensure the correct treatment of EM waves at the computational domain boundary for modeling microelectronic devices. 
Periodic boundary is implemented through the AMReX and WarpX infrastructure, where the EM waves leaving one face of the computational domain re-enter from the opposite face by exchanging guard-cell information handled by the AMReX code. 
To model a perfect conductor such that EM waves are reflected at the boundary, we have implemented the well-known perfect electric conductor (PEC) condition, by enforcing zero tangential electric field and normal magnetic field at the domain boundary.  
We have validated the PEC implementation by ensuring that the reflection coefficient is $-1$ for a simple test case with a sinusoid EM wave reflected from the infinite PEC boundary. 
%\MarginPar{@Jackie, should the reflection coefficient be 1 or -1?. JY --- for PEC, r=-1; for PMC, r=1.} 

The third boundary condition is an open boundary to model continuity of the medium and ensure EM waves cleanly exit the domain without re-entering or getting reflected. We achieve this by using a perfectly matched layer (PML) first developed by \cite{berenger1994perfectly,berenger1996three} boundary condition which damps out EM waves using a non-physical conductivity in the guard cells outside the domain boundary. 
Our solver leverages the two-step time-centered PML implementation, called ``PML-2SC'', implemented in WarpX by \cite{shapoval2019two}. However, the implementation only supported vacuum medium, and in our solver, we extended it to include macroscopic material properties such as $\sigma$, $\epsilon$ and $\mu$.
In the macroscopic PML-2SC implementation, field components $\Eb$ and $\Hb$ are updated using macroscopic Maxwell's equations that account for spatially varying material properties and then the updated field components are multiplied by the corresponding PML damping coefficients, similar to the approach described by \cite{shapoval2019two}. 
We have verified that the $\Eb$ and $\Hb$ fields are dissipated with $<0.1\%$ reflection for non-magnetic material aligned with the domain boundaries, wherein the relative permeability $\mu_r>=1$. 
Extending this boundary condition for magnetic materials using relative permeability self-consistently computed from the local magnetization, $\Mb$ and magnetic field, $\Hb$ will be implemented as future work.%\MarginPar{well not with magnetic; we may have to state this only works for non-magnetic? JY --- or say that we are working on PML touching LLG material?. the last line in this paragraph mentions that PML for magnetic materials is work in progress.}

%Typical boundary conditions required for microelectronics applications include periodic domain boundaries, perfectly conducting boundaries, and an open or continuous material boundary which allows for the electromagnetic waves to leave the domain without undergoing reflection.
%The periodicity of the electromagnetic wave is leveraged from the WarpX code, where, the AMReX geometry object is set to periodic and the guard-cell field exchange is handled accordingly. 

%In addition to periodic boundary conditions, we also support the use of perfect electric conductor (PEC) and perfectly matched layer (PML) conditions at the domain boundaries.
%The PEC implementation is straightforward, where, we hold values of tangential components of electric field, $\Eb$, and normal component of magnetic field, $\Hb$, fixed at zero at the domain boundaries. We have validated the PEC implementation by ensuring that the reflection co-efficient is one for a simple sinusoid electromagnetic wave. The PML implementation for materials is more involved, wherein, the fields at the domain boundary are damped., as we now describe.

%Our PML implementation is based on the PML-2SC variant of the two-step approach of \cite{shapoval2019two} which is based on the standard Berenger approach \cite{berenger1994perfectly,berenger1996three}

%State we do for $H$ equations.

%Varying material properties.

\section{Software}\label{sec:software}
We implement our code using AMReX developed by \cite{zhang2019amrex,zhang2020amrex}, which is a software framework developed and supported by the DOE Exascale Computing Project AMReX Co-Design Center.
AMReX contains many features for solving partial differential equations on structured grids; here we discuss relevant features for our present implementation, as well as future plans that will incorporate additional features.

AMReX manages data and operations on structured grids in a manner that can efficiently use the full range of computer architectures from laptops to manycore/GPU supercomputing architectures.
We divide the computational domain into non-overlapping grids, and each grid is assigned to an MPI rank.
AMReX uses an MPI+X fine-grained parallelization strategy, where X can be OpenMP (for multicore architectures), or CUDA (for GPU-based architectures).
Each of these strategies are implemented with the same front-end code using specialized looping structures within AMReX and the portability across various platforms is ensured by AMReX during compile-time.
Each MPI process applies computational kernels only to the data in grids that they own in the form of triply-nested loops (over each spatial dimension).
For pure MPI calculations, the loop is interpreted as a standard ``i/j/k'' loop.
For MPI+OpenMP calculations, the bounds of the loop is further subdivided over logical tiles, and each OpenMP thread loops over a particular tile.
For MPI+CUDA calculations, AMReX performs a kernel launch and offloads each data point to CUDA threads that in turn perform computations.
AMReX manages data movement by keeping data on the GPU devices as much as possible, avoiding costly communication between the host and device.
Thus, whenever possible, data movement to/from the host/GPU and also between MPI ranks is limited to ghost cell data exchanges, which occur a small number of times per time step.
In Section \ref{sec:Performance} we demonstrate the efficiency and scalability of our code using pure MPI, MPI+OpenMP, and MPI+CUDA simulations on NERSC systems. 
This analysis is made realized built-in AMReX profiling tools; however more in-depth analysis is possible with an extensive variety of compatible profilers such as CrayPat, IPM, and Nsight.
Data from the simulation can be efficiently written using a user-defined number of MPI ranks, to prevent overwhelming the system with simultaneous writes.%, as well as asynchronous I/O to background threads to allow for simultaneous continued computation.
Visualization can be performed with a number of publicly-available software packages, including Amrvis, VisIt, Paraview, and yt.

The code was originally created as a fork from the WarpX Maxwell solver in order to inherit the framework for the time-advancement, boundary conditions, and parsing of material properties and sources.
Since then, new field-update kernels have been added to support the features described in this paper such as spatially-varying material properties, the evolution of the magnetization and magnetic field, boundary condition support for PML and PEC, the ability to parse spatially-varying material properties ($\epsilon, \mu, M_s, \sigma, \gamma, \alpha$), and the ability to add external $\Eb$ and $\Hb$ waveforms (both hard and soft sources).
All codes are publicly available on the open-source github platform.
AMReX can be obtained at \url{https://github.com/AMReX-Codes/amrex} and
the microelectronics code can be obtained at \url{https://github.com/ECP-WarpX/artemis}.

\subsection{Code Performance}\label{sec:Performance}
To assess the scaling and performance of our code on HPC systems we perform tests on the NERSC supercomputer.
We consider both the Haswell and GPU partitions on the Cori system.
The Haswell partition consists of 2,388 nodes; each node contains an Intel Xeon ``Haswell'' Processor with 32 physical cores and 128GB memory.
The GPU partition is a test-bed representative of the forthcoming Perlmutter system consisting of 18 nodes; each node contains a 20-core Intel Xeon Gold ``Skylake'' processor with 372GB memory and 8 NVIDIA Tesla V100 ``Volta'' GPUs each with 16GB memory.
We perform weak and strong scaling tests and then compare kernel computation and communication times across the different architectures using pure MPI, MPI+OpenMP, and MPI+CUDA.

To perform weak scaling studies we use the magnetically tunable filter case described in Sec.~\ref{sec:MTF}.
We perform three sets of simulations to demonstrate performance portability across heterogeneous computer architectures on the NERSC Cori systems.
The first set of simulations employ the pure MPI paradigm on the Haswell partition with a maximum of 32 MPI processes per node.
For the second set of simulations, we use a hybrid MPI+OpenMP approach on the Haswell partition, where we use up to 8 MPI processes per node, and 4 OpenMP threads per MPI process. %\MarginPar{I forget why we chose 4 OpenMP threads?  AJN: no particular reason; it was a reasonable number to try}
The third set of simulations employ the MPI+CUDA parallelization strategy on the GPU partition of Cori, which has a maximum of 8 NVIDIA V100 GPUs, and we use one MPI processor per GPU.
Note that a node on the GPU partition contains 20 physical cores on the host, yet we only use a maximum of 8 cores (one per GPU).

The baseline case for all weak scaling tests performed with different parallelization strategies has a grid size of 1024 $\times$ 4 $\times$ 512 cells (2,097,152 total).
For this base case we use 1/8th of a node for each set of simulations, i.e.~4 MPI ranks on a Haswell partition node, 1 MPI rank $\times$ 4 OpenMP threads on a Haswell partition node, and 1 MPI rank + 1 GPU on a GPU partition node.
We note that this simulation uses roughly 4GB memory, which is 25\% of the memory available on 1/8th of a Haswell partition node and also 25\% of the memory available on 1/8th of the Cori GPU node.
%\MarginPar{the concept of grid -- I think is special for AMReX in that -- I think -- typically people think of grid as the whole simulation. But we think of grid and boxes in the same way. Should we say, the domain is divided into boxes or sub-grids, and we assign one box per MPI rank, .... JY: How about the added red sentence?}
The entire spatial domain is divided into grids (commonly called sub-grids in other EM simulation codes), and we assign 1 grid per MPI process; thus the 4 MPI-rank run uses four 1024 $\times$ 4 $\times$ 128 grids and the 1 MPI-rank runs use one grid spanning the domain.
For the OpenMP runs we use a tile size of 1024 $\times$ 4 $\times$ 8 to subdivide each grid.
For the weak-scaling study, we increase the domain size in the $z$ direction consistent with the increase in the core count to maintain a constant amount of computational work per core.

\begin{figure}[tb]
\centering
\includegraphics[width=0.5\textwidth]{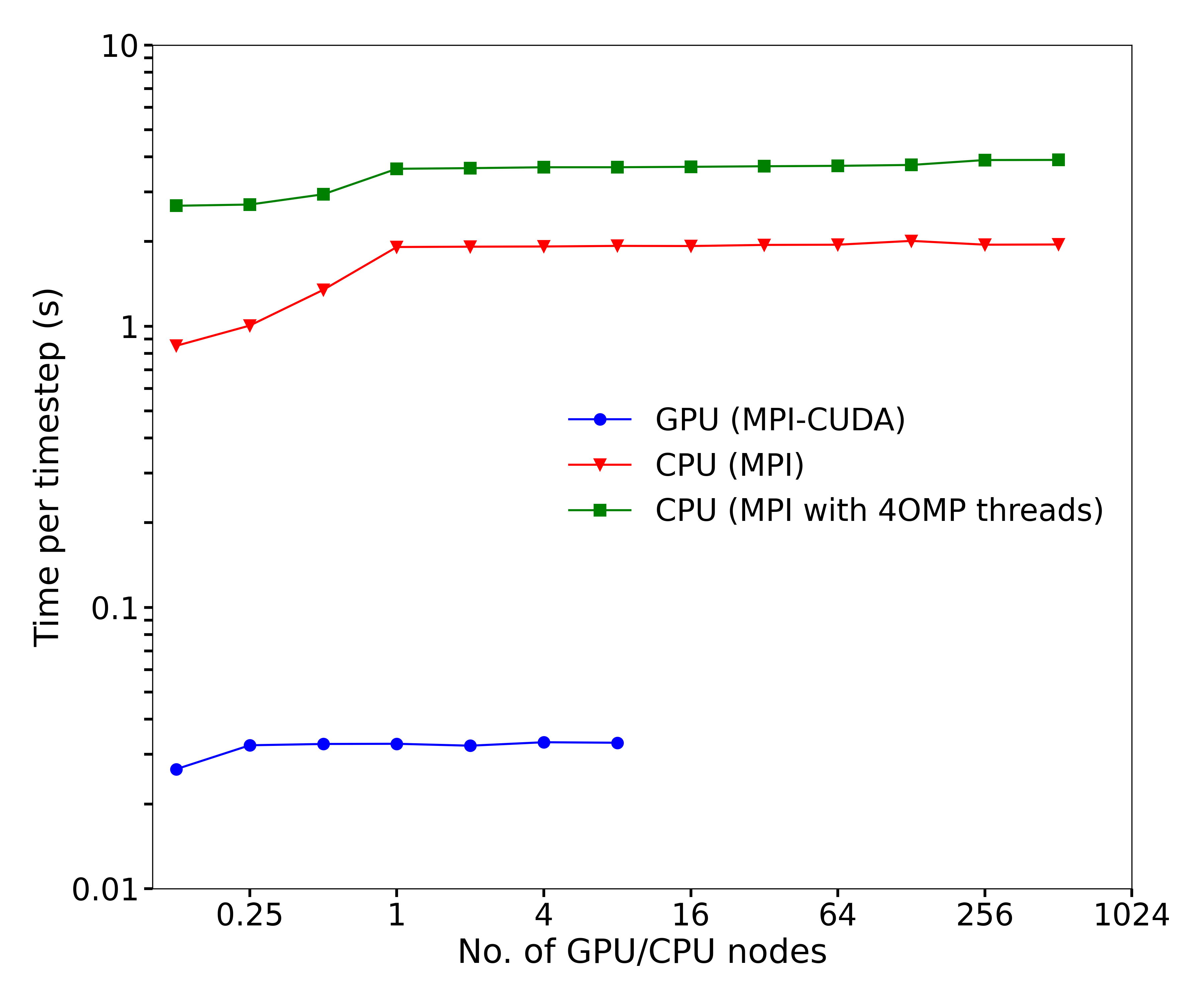}
\caption{Weak scaling for pure MPI, MPI+OpenMP, and MPI+CUDA on the NERSC Cori Haswell and GPU partitions.  Note that one Haswell partition node contains 32 physical cores, and one GPU partition node contains 8 GPUs.
The weak scaling is nearly perfect for the Haswell partition simulations past the 1 node threshold up to 20\% of the entire system.  The weak scaling is also nearly perfect for the GPU partition simulations past the 1/4-node threshold up to 50\% of the testbed.  Also, using 1 node or more the GPU simulations run 59x and 112x faster than the pure MPI and MPI+OpenMP simulations on a node-by-node comparison.}
\label{fig:WeakScaling}
\end{figure}

Fig. \ref{fig:WeakScaling} shows weak scaling results for the three sets of simulations, indicating the simulation time per time step as a function of the number of nodes.
We omit the timing associated with initialization, including memory allocation and data transfer on the host and GPU, as these occur only once at the beginning of the simulation.
We note that we performed tests on the Haswell partition up to 512 nodes (16,384 cores), which is roughly 20\% of the entire system.  We performed tests on the GPU partition using up to 8 nodes (64 GPUs), which is 50\% of the testbed.
For the 1/8th node baseline runs, the GPU simulations performed 32 times faster than the pure MPI simulations on the Haswell partition, and 100 times faster than the MPI+OpenMP simulations on the Haswell partition.
At a glance, we see that for the 1-node through 8-node runs, the GPU simulations performed approximately 59x faster than the pure MPI simulations on the Haswell partition, and approximately 112x faster than the MPI+OpenMP simulations on the Haswell partition.
We attribute the factor of two slowdown with MPI+OpenMP compared pure-MPI to the overhead required to spawn the OpenMP threads.
The Haswell partition simulations achieve nearly perfect scaling up to 512 nodes beyond the 1-node threshold. The reason is that as we increase the number of processors within the node, the MPI communication-time keeps increasing until the node is saturated and then the computation-to-communication ratio remains the same across all nodes.
The GPU partition simulations achieve nearly perfect scaling up to 8 nodes beyond the 1/4-node threshold, i.e., 2 MPI-GPU processors. 
The increase in simulation time from 1-GPU to 2-GPU simulation is mainly dominated by the communication time across their respective MPI hosts.
%We note that on the Haswell partition, the scaling does not become flat until the 1-node threshold is achieved, probably due to increased intra-node communication.
In this paper we are not using dynamic load balancing and assign an equal number of grid points to all processors using a fixed grid structure.
In the future, as we move to adaptively refined simulations, dynamic load balancing will become more important.
One particular code optimization we may explore is the use of grid structures where magnetic regions contain a larger number of smaller grids; the computational requirements for the evolution of $\Mb$ are much more demanding than $\Eb$ and $\Hb$ so we would be focusing computational resources in these regions.
The communication exchange in our algorithm is relatively small, where ghost cell exchanges occur once after each variable advancement routine.
There are no communication-intensive linear solvers in our implementation.
Thus, we expect this scaling trend to continue on the forthcoming Perlmutter system since we have demonstrated nearly perfect scaling with 16,384 MPI processes on the Haswell partition.

\begin{figure}[t]
\centering
    \subfloat[\centering] {\includegraphics[width=0.45\textwidth]{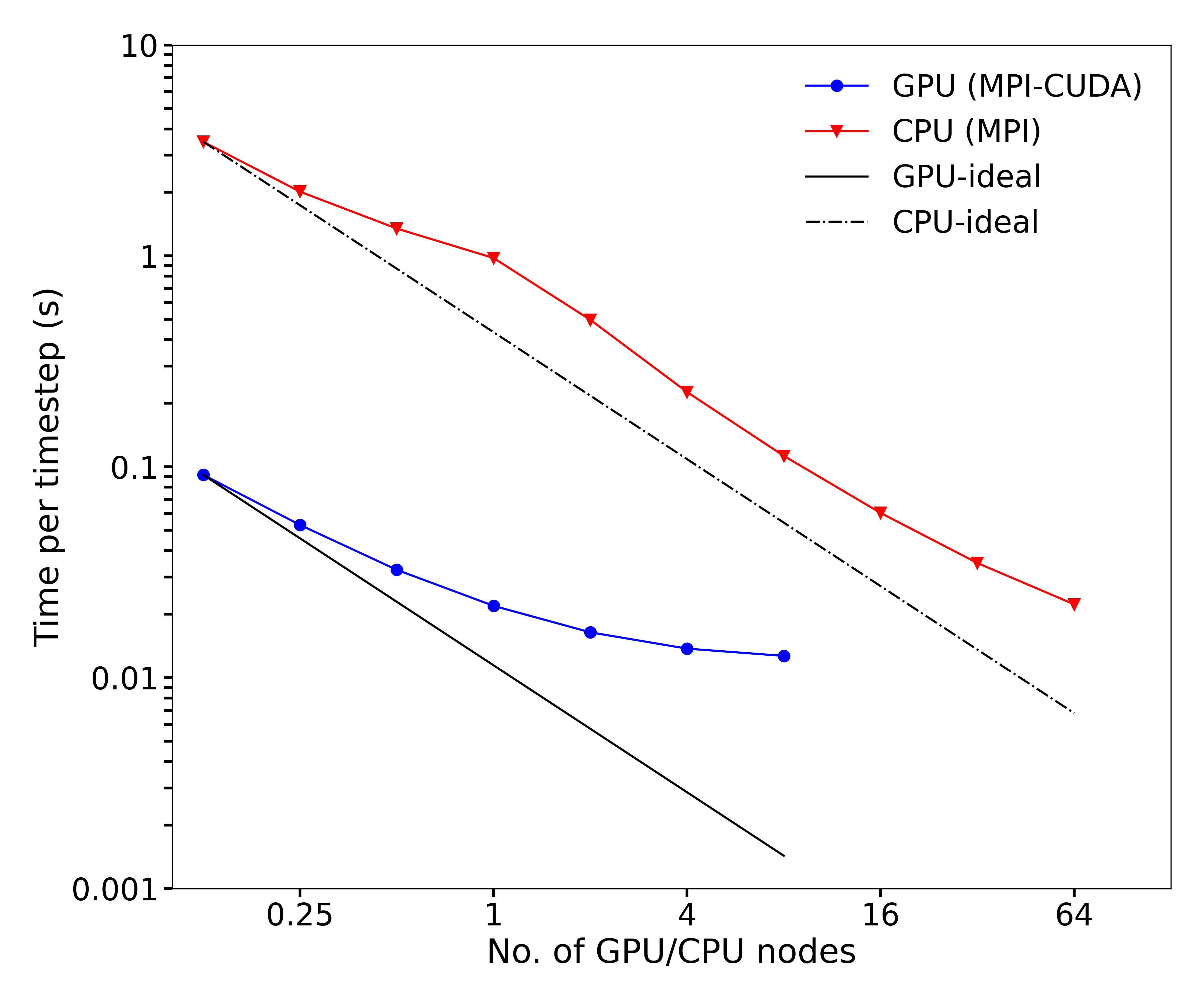}} \\
    \subfloat[\centering]{\includegraphics[width=0.45\textwidth]{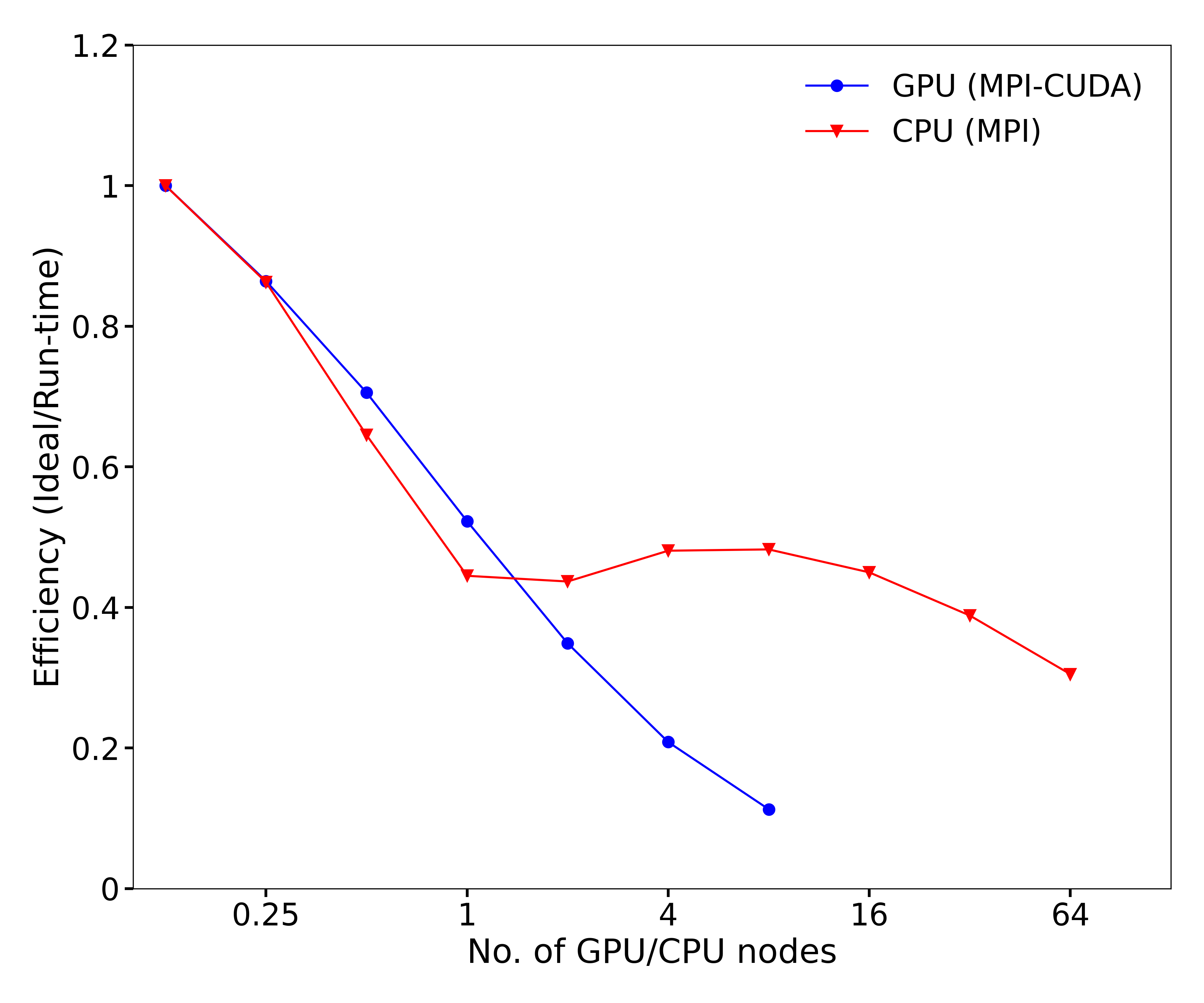}}
    \caption{(a) Strong scaling performance and strong scaling efficiency on the NERSC Cori Haswell and GPU partitions.  The departure from ideal scaling is more pronounced on the GPU partition runs, as there is not enough computational work per GPU to make the most effective use of the system.}
    \label{fig:StrongScaling}
\end{figure}

For our strong scaling tests, we chose a problem with $1024 \times 4 \times 2048$ cells for all the simulations, as it uses nearly all the available memory on 1/8th of the Haswell and GPU partition node.
We performed two sets of simulations, pure MPI and MPI+CUDA.
In both sets we hold a constant number of grid points in the domain, and divide the domain into a larger number of smaller grids such that each MPI process is assigned one grid.
In Fig. \ref{fig:StrongScaling}, we show strong scaling and the corresponding strong scaling efficiency, the latter defined as the ratio of runtime to ideal scaling runtime from the 1/8th node cases.
%\MarginPar{remove green curves in figure \ref{fig:StrongScaling}}
The departure from ideal strong scaling occurs simply because there is not enough work per MPI process compared to the non-compute communication overhead as a larger number of smaller grids are used, which is an effect seen most noticeably in the GPU runs.
As the compute work per processor decreases, the computational overhead, notably communication, takes a larger percentage of runtime.
As expected, the GPU runs outperform the MPI runs by an increasingly larger margin as the compute work per MPI process increases.

\begin{table*}[t]
\small\sf\centering
\caption{Profiling output for single-node simulations for pure MPI, MPI+OpenMP, and MPI+CUDA over 10 time steps. We indicate percent of runtime for each routine in the first three columns, and the speedup of the GPU runs compared to the non-GPU runs in the last two columns.}
\begin{tabular}{lccccc}
\toprule
& & & & Speedup & Speedup \\
Kernel name & pure MPI (s) & MPI+OpenMP (s) & MPI+CUDA (s) & pure MPI & MPI+OpenMP\\
\midrule
\texttt{Evolve HM} & 8.38 (54\%)  &  21.05 (71\%)  & 0.1 (38\%)    & 84x  & 211x \\
\texttt{Evolve E}      & 1.13 (7.3\%) &  1.79  (6.1\%) & 0.007 (2.6\%) & 161x & 256x \\
\texttt{Communication} & 6.01 (39\%)  &  6.73  (23\%)  & 0.158 (60\%)  & 38x  & 43x \\
\midrule
Total Time    & 15.52        &  29.57         & 0.265         & 59x  & 112x \\
\bottomrule
\end{tabular}\\[10pt]
\label{tab:profiler}
\end{table*}

We detail the timing of various routines for the three sets of simulations (on one single node) in Tab. \ref{tab:profiler}.
We see that each individual kernel, as well as the communication overhead, are significantly faster on the GPU partition, such that the overall speedup is 59x and 122x compared to the pure MPI and MPI+OpenMP simulations.
The speedup for the computational kernels ``Evolve HM'' and ``Evolve E'' is significant, ranging from 84x up to 256x.
We note that unlike the CPU-simulations, communication dominates the GPU simulations requiring 60\% of the total runtime. 
Nevertheless the time spent in communication in the GPU simulations is still 38x and 43x smaller compared to the MPI and MPI+OpenMP run. 

\section{Validation}\label{sec:validation}

\subsection{Convergence Test}
We performed a coupled LLG-Maxwell test-case to demonstrate the order of accuracy of the numerical scheme previously described in Sec.\ref{sec:numerical} with spatially varying material properties. 
In our test problem, we consider a fully periodic domain with $L=0.1$~m (all units in this paper are MKS unless noted otherwise) on each side, centered about the origin.
We initialize a three-dimensional EM signal, such that at $t=0$,
\begin{equation}\label{eq:convergeE}
\begin{split}
E_x = 10^6 \cos{(20\pi z)},\\
E_y = 10^6 \cos{(20\pi x)},\\
E_z = 10^6 \cos{(20\pi y)},
\end{split}
\end{equation}
\begin{equation}\label{eq:convergeH}
\begin{split}
H_x = \frac{10^6}{120\pi}\cos{(20\pi y)}, \\
H_y = \frac{10^6}{120\pi}\cos{(20\pi z)}, \\
H_z = \frac{10^6}{120\pi}\cos{(20\pi x)}.
\end{split}
\end{equation}
Note each of the electric and magnetic field components in (\ref{eq:convergeE}) and (\ref{eq:convergeH}) satisfy the relation $E_p/H_p = 120\pi = \sqrt{\mu_0/\epsilon_0}$, with the ratio being the intrinsic impedance of vacuum. 
%\MarginPar{The $120pi$ comes from some combination of mu0, c, eps0, or something like that. It would help to mention that.. so that others can reproduce it, if required.}
Defining $r$ as the distance from the origin (center of the domain), the material properties have the following spatially-varying profiles,
\begin{eqnarray}
\sigma &=& 0.1e^{-1000r^2}, \\
\epsilon &=& \epsilon_0 (1 + e^{-1000r^2}), \\
M_s &=& 10^4 (1 + e^{-1000r^2}), \\
\alpha &=& 0.5 (1 + e^{-1000r^2}), \\
\gamma &=& -1.759 \times 10^{11} (1 + e^{-1000r^2}).
\end{eqnarray}
We also set $M_x = M_s$ and $M_y = M_z = 0$ at the beginning of the simulation.
Note that since $M_s>0$, only vacuum permeability is used to update the fields as discussed in Sec. \ref{sec:model}.
In this particular test, we do not include any additional bias so that $\Heff=\Hb$.
We perform three simulations with $32^3$, $64^3$, and $128^3$ grid size, such that, we increasingly refine the cell-size of each dimension by a factor of two. 
We call these test cases as coarse (c), medium (m), and fine (f), respectively.
To compare the results from these simulations at the same physical time, a CFL=0.9 is used to define the timestep based on the speed of light in vacuum and the simulations are run for 5, 10, and 20 timesteps, respectively, to reach the same physical time. 
The convergence rate is defined as the base-2 log of the ratio of errors between the coarse-medium, $E_c^m$, and medium-fine solutions, $E_m^f$.
\begin{equation}
{\rm Rate} = \log_2\left(\frac{E_c^m }{E_m^f}\right)
\end{equation}
 $E_c^m$ and $E_m^f$ are obtained by computing the $L^1$ norm between the coarser and the finer solution of a field, $\phi$, averaged/interpolated to the same grid points as the former. 
 Note that, for face-centered data, the solutions from the finer simulation is re-constructed on the coarser grid by averaging the 4 fine face-centered values overlying the coarser grid. Similarly, for the edge-centered data, the finer solutions are re-constructed on the coarser grid by averaging the two fine-edge values overlying the coarse-edge.
 Errors $E_c^m$ and $E_m^f$ are defined as,
\begin{equation}
E_c^m = \frac{1}{Nc_{\rm pts}} \sum_{i,j,k} |\phi_{m} - \phi_{c}| ;
E_m^f = \frac{1}{Nm_{\rm pts}} \sum_{i,j,k} |\phi_{f} - \phi_{m}| 
\end{equation}
where, $\phi_c$, $\phi_m$, and $\phi_f$ are the fields obtained from the coarse, medium, and fine solutions, and $Nc_{\rm pts}$ and  $Nm_{\rm pts}$ are total number of grid points used for the coarse and medium simulations, respectively. 
As shown in Tab. \ref{tab:convergence}, we obtain clean second-order convergence for all variables. Even though we only report convergence in $L^1$, we obtain nearly identical convergence rates in $L^0$ and $L^2$.
We have also performed similar tests using the first-order magnetization coupling scheme described in Sec.~\ref{sec:first-order}.  
As expected, the convergence rates for the $\Mb$ components dropped to first order, and the rates for the $\Eb$ and $\Hb$ fields reduced to as low as 1.74 due to coupling with the first-order $\Mb$ field.
%was more erratic, with convergence rates ranging from 1.74 to 2.35. 
%\MarginPar{do we know why? may be because of the scalar form, and Maxwell is always second order??, but the first-order Mfield coupled with H reducing its order?  }

\begin{table}[h]
\small\sf\centering
\caption{Convergence rates in the $L^1$ norm for all field variables.}
\begin{tabular}{cccc}
\toprule
Variable & $E_{32}^{64}$ & $E_{64}^{128}$ & Rate\\
\midrule
\texttt{$E_x$} & 1819.4 & 455.23 & 2.00 \\
\texttt{$E_y$} & 1820.5 & 455.04 & 2.00 \\
\texttt{$E_z$} & 1882.8 & 470.47 & 2.00 \\
\texttt{$H_x$} & 9.6659 & 2.4359 & 1.99 \\
\texttt{$H_y$} & 9.4752 & 2.3788 & 1.99 \\
\texttt{$H_z$} & 9.4590 & 2.3607 & 2.00 \\
\texttt{$M_x$} & 5.1343 & 1.2870 & 2.00 \\
\texttt{$M_y$} & 4.7851 & 1.1989 & 2.00 \\
\texttt{$M_z$} & 4.9642 & 1.2429 & 2.00 \\
\bottomrule
\end{tabular}\\[10pt]
\label{tab:convergence}
\end{table}

%We use a CFL of 0.9 based on the speed of light in vacuum to define the time step and run the simulations 5, 10, and 20 time steps, respectively, to the same physical time. 
%The error at a given resolution with cell-size, $dx$, is obtained by first coarsening the simulation result obtained with a factor of two smaller cell-size, $dx/2$, then computing norms on the difference between the data. 
%To compute the convergence rates, we first define the error in a coarsened field variable, $\phi_c$, as the $L^1$ norm of a coarsened version of the next-finer solution, $\phi_f$ subtracted from the coarse solution,
%The error
%\begin{equation}
%E_{\rm coarser}^{\rm finer} = \frac{1}{N_{\rm pts}} \sum_{i,j,k} |\phi_{\rm finer} - \phi_{\rm coarser}|
%\end{equation}
%where the sum is taken over all coarse grid points and $N_{\rm pts}$ is the total number of coarse grid points.
%For face-data (components of $\Hb$), $\phi_f$ is constructed by averaging the 4 fine face values overlying the coarse face.
%For edge-data (components of $\Eb$ and $\Mb$), $\phi_f$ is constructed by averaging the 2 fine edge vlues overlying the coarse edge.
%The convergence rate is defined as the base-2 log of the ratio of errors between the coarse-medium and medium-fine solutions.
%\begin{equation}
%{\rm Rate} = \log_2\left(\frac{E_c^m }{E_m^f}\right)
%\end{equation}
%As shown in Tab. \ref{tab:convergence}, we obtain clean second-order convergence in all variables. Note that we obtain nearly identical second-order convergence rates for the $L^0$ and $L^2$ norms as well.
We only report convergence in $L^1$, but we note that we obtain nearly identical convergence rates in $L^0$ and $L^2$.
We also performed similar tests using the first-order magnetization coupling scheme described in Sec.~\ref{sec:first-order}.  As expected, the convergence rates for the $\Mb$ components dropped to first order, and the rates for the $\Eb$ and $\Hb$ fields was more erratic, with convergence rates ranging from 1.74 to 2.35. 
%\MarginPar{do we know why? may be of the scalar form? }

\subsection{Practical Device Applications}

We validate our code by simulating the characteristics of two different practical waveguide devices.
We perform two validation tests, one to validate the core macroscopic Maxwell module, and second to validate the coupled Maxwell-LLG implementation.
In the first case, the standard X-band waveguide is filled with air (see Fig. \ref{fig:WG_empty}) to demonstrate EM wave propagation in the presence of known boundary conditions.
In the second case shown in Fig. \ref{fig:WG_ferrite}, we model a magnetically tunable filter realized by a ferrite loaded X-band waveguide structure. 
The magnetic spins interact with the elliptically polarized EM wave in the waveguide, and thus the wave exhibits exotic behaviors such as non-reciprocity and high attenuation around the ferromagnetic resonance (FMR) frequency. 
Thus, this test-case serves as a good validation for the coupled Maxwell-LLG algorithm. 
%\MarginPar{Are there references for these examples?}

\subsubsection{Air-filled Waveguide}
\label{sec:AFW}

\begin{figure}
\centering
    \subfloat[\centering] {\includegraphics[width=0.3\textwidth]{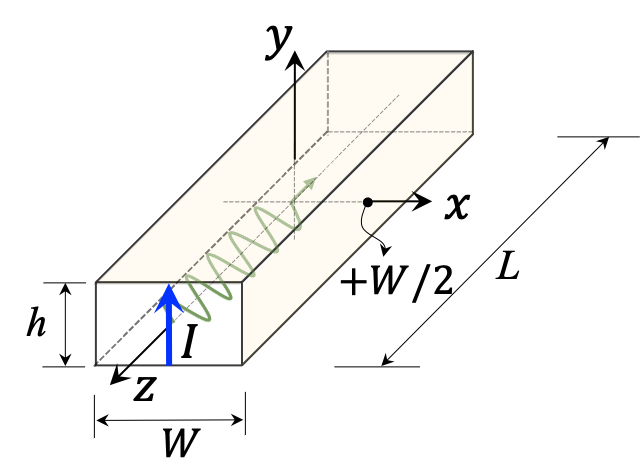} \label{fig:WG_empty}} \\
    \subfloat[\centering]
    {\includegraphics[width=0.24\textwidth]{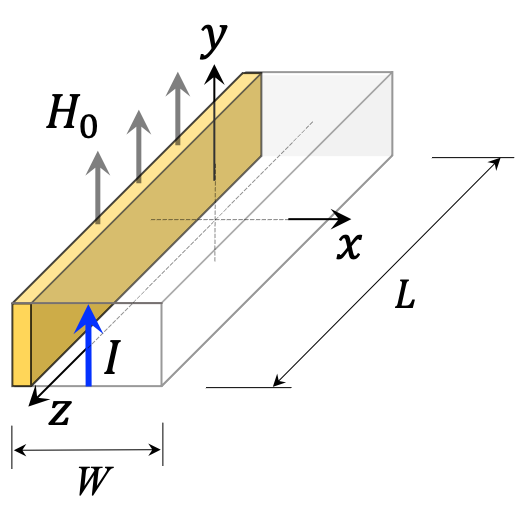} \label{fig:WG_ferrite}}
    \caption{(a) Air-filled waveguide. (b) Magnetically tunable filter, realized by the same waveguide structure in Fig. \ref{fig:WG_empty}, with an inserted ferrite slab.}
    \label{fig:three graphs}
\end{figure}

The air-filled rectangular waveguide described by \cite{Pozar2012} is used to validate the core macroscopic EM module in our solver.
As illustrated in Fig. \ref{fig:WG_empty}, the width of the waveguide is $W = 14.95$ mm, and the height is $h = 10.16$ mm. 
We apply PEC boundary condition at the four side walls (in $x$ and $y$) of the waveguide.
PEC is also applied at one end in the longitudinal direction (in $z$), reducing the simulation volume by a factor of 2. 
This type of boundary conditions is also referred to as anti-symmetric boundary conditions in other numerical frameworks. 
In this case, we apply this PEC boundary condition at $z=+L/2$, where $L$ is the length of the domain and equal to $500$ mm.
At the same location, we superimpose an electric current source $I$ in addition to the PEC boundary, which excites the EM wave propagating in the $-z$ direction.
Following conventions, the waveguide operates at its fundamental mode, or in the other words, the $\mathrm{TE}_{10}$ mode, with a cutoff frequency of $f_c = c/2W = 10.03 $ GHz, where $c$ is the speed of light in vacuum. 
To capture the cutoff knee point, the current excitation $I$ takes a modified Gaussian pulse form, as shown below:
%\MarginPar{How does $I$ translate to $E$ in the external source - and is this hard or soft?  This doesn't use $\Jb_{\rm src}$, does it?}
\begin{equation}
    I = \mathrm{exp}\left[\frac{-(t-3T_p)^2}{2T_p^2}\right] \mathrm{cos}\left(\omega _0 t\right) \mathrm{cos}\left(\frac{2\pi x}{W /2} \right), \label{eq:gaulssian} \\
\end{equation}
In (\ref{eq:gaulssian}), $\omega _0 = 2\pi f_0$ sets the central frequency of the pulse, with the value of $f_0$ being $10.5 \mathrm{GHz}$. 
$T_p$ is the time width of the pulse, which is set to be one period of the excitation frequency, i.e. $T_p=0.095238$ ns. 
Note that we apply a shape function $\mathrm{cos}(\frac{2\pi x}{W /2})$ for the purpose of mode matching to the $\mathrm{TE}_{10}$ mode. 
%\sout{According to the boundary relation between the magnetic field $H$ and the surface current $J_S$ ($\Delta H_{\rm{tangential}} = J_S$, and the field inside PEC is zero ), the current excitation is directly added on top of the $H$ field at the PEC.}
The current excitation is directly added on top of the $H$ field at the PEC, such that it satisfies the boundary condition,  $\Delta H_{\rm{tangential}} = J_S$ at $z=250 \rm{mm}$, where $J_s$ represents the surface current density in the unit of A/m.
%\sout{The length of the waveguide is $L = 500 \mathrm{mm}$, which is about 17 times of the EM wavelength in vacuum.}
In this simulation, we use $512 \times 4 \times 512$ grid points.
The CFL factor is 0.8, leading to a time step of $\Delta t =77.87961902$ fs.
We run the simulation to 80,000 time steps, which corresponds to a total of $\sim$65 EM wave cycles.

\begin{figure}[tb]
\centering
\includegraphics[width=0.5\textwidth]{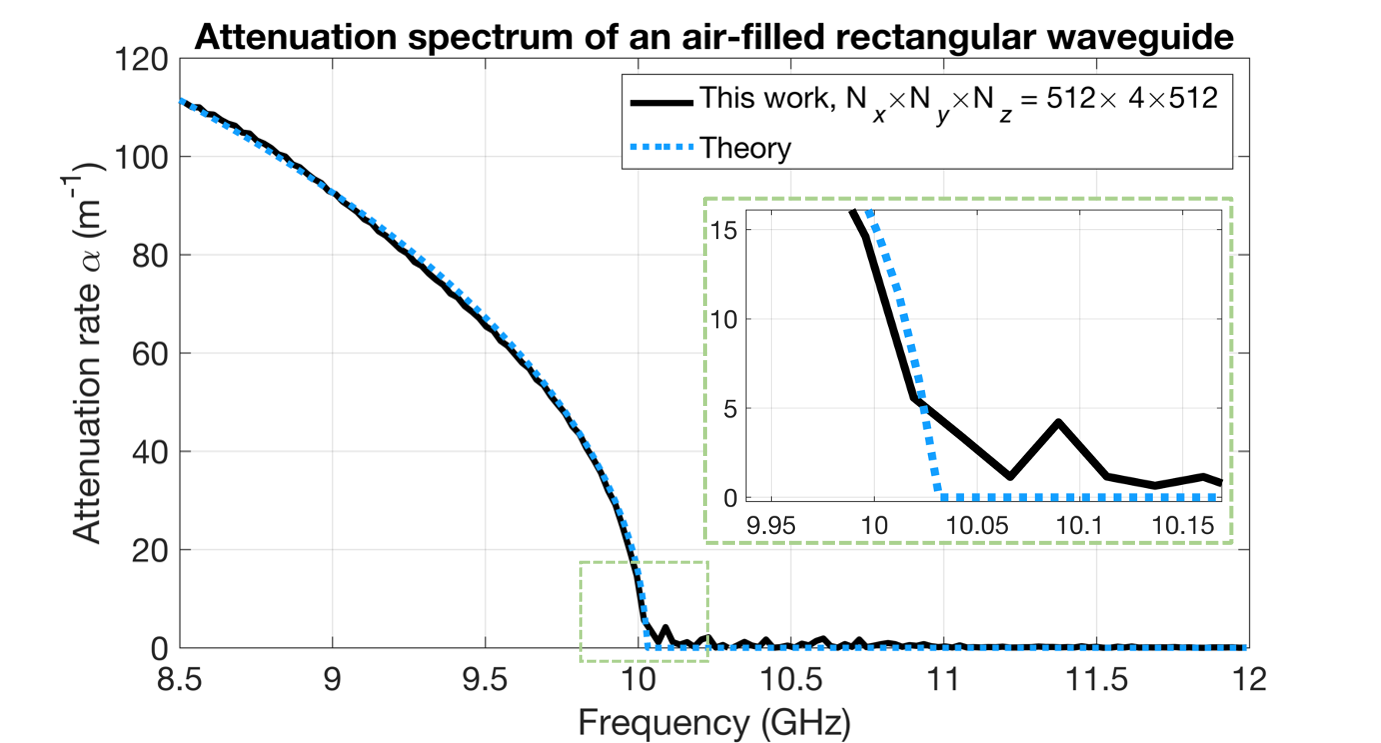}
\caption{Comparison of attenuation spectrum of the air-filled waveguide obtained from computations with theory.}
\label{fig:alpha_empty}
\end{figure}

After the EM fields reach steady-state, we compute fast Fourier transform on the time-domain electric field  at the central longitudinal line of the waveguide, i.e. $(x=0, y=0, -L/2 \leq z \leq +L/2)$. 
The attenuation constant of the electric field is derived from the extracted damping speed of the field magnitude, with the signal-processing code, ESPRIT, developed by \cite{esprit1,esprit2}.
Fig. \ref{fig:alpha_empty} shows both the theoretical and simulated attenuation spectrum of the air-filled waveguide. 
The theoretical attenuation rate $\alpha$ is defined as
\begin{equation}
\alpha = \mathrm{imag}\left[\sqrt{\omega ^2 \mu _0 \epsilon _0 - (\pi / W)^2}\right]
\end{equation}
The theoretical cutoff frequency, $f_{c1}$, of this waveguide structure is $f_{c1} = c/(2W) = 10.03 \mathrm{GHz}$.
It can be observed that the $\alpha$ follows the hyperbolic shape before the cutoff frequency $f_{c1}$. 
Beyond the cutoff frequency, $\alpha$ becomes zero, indicating that the wave propagates along the waveguide with no attenuation. 
Both the curve shape and the cutoff frequency agree well the theoretical prediction.

\subsubsection{Magnetically tunable filter}
\label{sec:MTF}
To validate the interaction between the magnetic spins and the EM wave in the coupled Maxwell/LLG framework, a thin ferrite slab is inserted in the longitudinal direction of the waveguide, as shown in Fig. \ref{fig:WG_ferrite}. 
The ferrite material is yttrium iron garnet (YIG), which is widely used in RF front-ends and devices, due to its narrow magnetic linewidth $\Delta H$.
The dimensions and setup of this YIG-inserted waveguide are the same as that for the air-filled waveguide shown in Fig. \ref{fig:WG_empty}.
In our simulation, the saturation magnetization is $4\pi M_S = 1750$ Gauss, or $1.3926e5$ A/m.
A DC magnetic bias $H_0 = 2890~\mathrm{Oersted} = 2.2998e5~\mathrm{A/m}$ is applied along the in-plane direction of the ferrite slab and the tangential direction of the waveguide, therefore, $H_0$ is in the $y$ direction.
According to Kittel's equation, the FMR frequency of the thin slab is 
\begin{equation}
    f_\mathrm{FMR} = 2.8~\mathrm{MHz} \times \sqrt{H_0 (H_0 + M_S)}. \label{eq:kittel}
\end{equation}
%\MarginPar{generally, we dont have units show up in the middle of an expression. Do we need the 2.8 MHz there? Can we use a variable for it and explain the the variable = 2.8 MHz? JY: I'd say this is the convention that people use in Kittel equation, I know, weird...}
Note in (\ref{eq:kittel}), $H_0$ and $M_S$ are both in units of Oersteds. Therefore, the theoretical FMR frequency is 10.25 GHz.
%\MarginPar{is Oersteds MKS? is not, we have a statement in Convergence test where we say all units in this paper are in MKS. :) JY: Good point. I added "unless noted otherwise" at the end of that sentence}
%\MarginPar{Comment why theory, experiment, and our result show more like 10.4GHz? JY --- added in Line 759}
The damping factor of YIG is $\alpha = 0.0051$, corresponding to an FMR linewidth of $35~\mathrm{Oersted}$ at X band, and its relative permittivity is $\epsilon_r = 13$.
%\sout{The relative permittivity of YIG is $\epsilon_r = 13$.}
The thickness of the YIG slab is 0.45 mm, with one of the faces touching the PEC wall of the waveguide.
%\textcolor{blue}{The dimensions and setup of the \textcolor{red}{YIG} waveguide \textcolor{red}{are the same as that for the air-filled waveguide shown in Fig. \ref{fig:WG_empty}.} \sout{stay the same to the case of Fig. \ref{fig:WG_empty}.}}

\begin{figure}[tb]
\setlength{\fboxsep}{0pt}%
\setlength{\fboxrule}{0pt}%
\begin{center}
\includegraphics[width=0.51\textwidth]{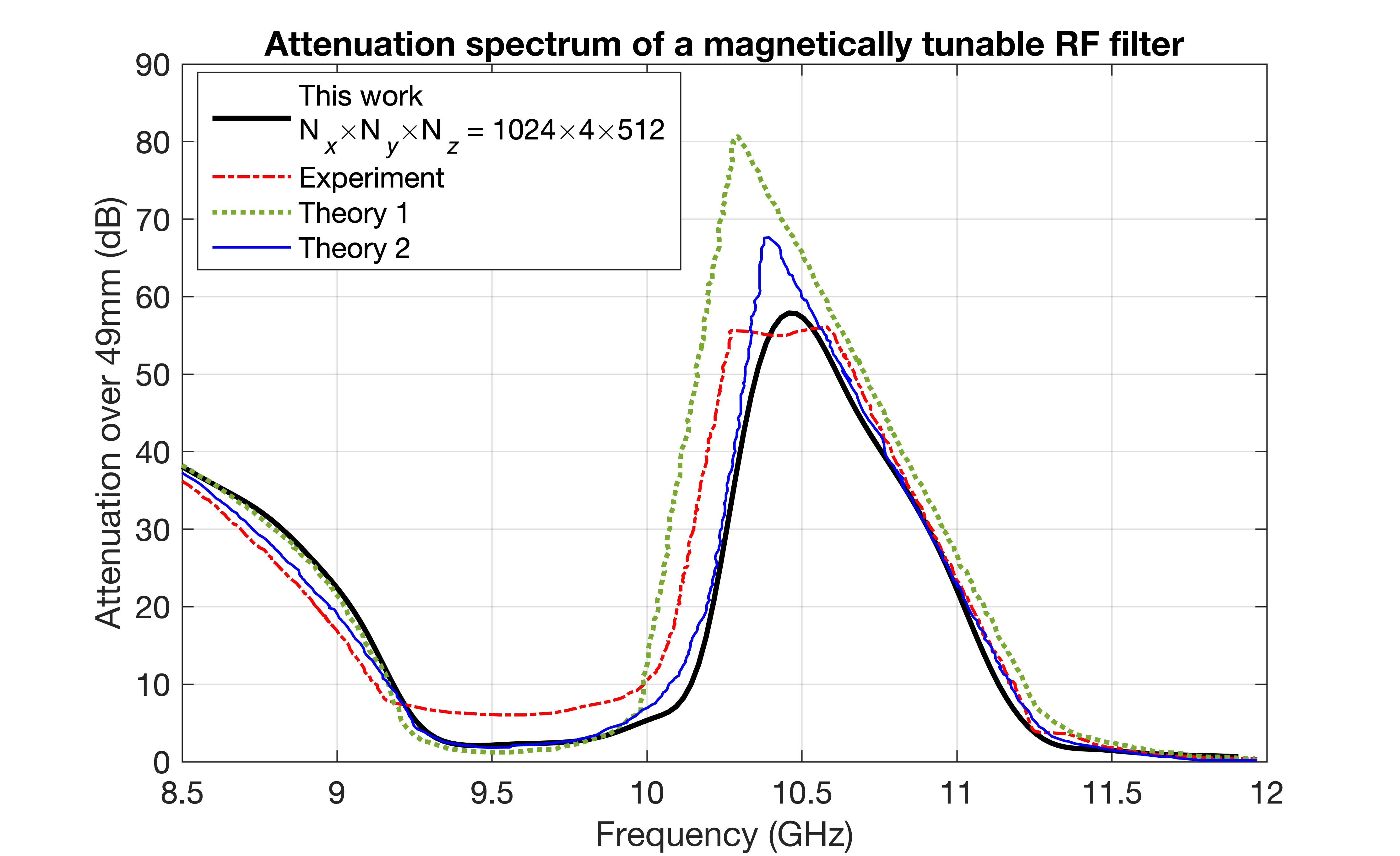}
\end{center}
\caption{Comparison of computation result to experimental result and theoretical predictions for the attenuation spectrum of the magnetically tunable filter.  The key features of the attenuation spectrum obtained from the simulations agree with experiments and theory.}
\label{fig:alpha_YIG}
\end{figure}

We use the same post processing technique to extract the attenuation spectrum $\alpha(f)$ as that described in Sec.~\ref{sec:AFW}.
Fig. \ref{fig:alpha_YIG} shows the comparison of the attenuation spectrum obtained from experimental tests reported by \cite{gardiol1971}, theoretical analyses reported by \cite{gardiol1971,uher1987}, and the simulation of this work.
%Our simulated result matches all the qualitative features of both theory and experiment.
The key features of the attenuation spectrum, such as waveguide cut-off frequency and FMR frequency, obtained from the simulations agree with experiments and theory.
%\sout{The key features of the attenuation spectrum obtained from the simulations agree with experiments and theory.
%In particular, the waveguide cutoff feature is clearly seen from Fig. \ref{fig:alpha_YIG} for all curves, elaborated as following.}
$\alpha(f)$ shows a hyperbolic shape below 9.2 GHz, which is the cutoff frequency $f_{c2}$. 
Note that $f_{c2}$ is different from $f_{c1}$ by 0.83 GHz, primarily due to the large relative permittivity $\epsilon_r = 13$ present in the YIG slab. 
Moreover, the largest attenuation rate of the EM wave is observed around 10.45 GHz. 
%The discrepancy between the FMR frequencies predicted by the Kittel's equation and simulation/experiment can be attributed to the assumption of infinite planar dimension of the thin film used in the theory compared to finite-size thin films used in simulation and experiment.
%Due to similar reasons, our results do not recover as sharp of a peak at FMR as theory, and is in better agreement with experimental measurements.
Tab. \ref{tab:alpha_YIG} shows a quantitative comparison of the cut-off frequency, FMR frequency, and peak attenuation obtained from the spectrum-curves shown in Fig. \ref{fig:alpha_YIG}. 
The cut-off frequency from the simulation is within 1.2\%,  0.35\% and 0.25\% discrepancies compared to experiment,  theory 1 and theory 2, respectively.
Similarly, the FMR frequency from the simulation is within 0.21\%, 1.6\% and 0.56\% discrepancies compared to experiment, theory 1 and theory 2, respectively.
However, the peak attenuation varies by 5.2\%, 28\% and 14\% compared to experiment, theory 1 and theory 2, respectively.
%As mentioned earlier, the difference between simulation and theory is higher because of the infinite material dimension assumed in the theoretical analysis.
The discrepancy between the FMR frequencies predicted by the Kittel's equation and simulation/experiment can be attributed to the assumption of infinite planar dimension of the thin film used in the theories compared to finite-size thin films used in simulation and experiment.
Due to similar reasons, our results do not recover as sharp of a peak at FMR as theory, and is in better agreement with experimental measurements.
%Our simulations agree with experiments and theory within 5\% and 10\% respectively.
%\MarginPar{Does this suggest something isn't quite right with the theory?  As in there needs to be some kind of correction factor?}
%The simulated result of this work matches better to the experimental result with closer values of $\alpha _\mathrm{peak}$ and $f_\mathrm{FMR}$.
%The values of $f_\mathrm{c2}$ match better among the theories and our simulation.\MarginPar{explain why? no clue...}

\begin{table*}[t]
\small\sf\centering
\caption{Key features of all attenuation spectra curves, $\alpha(f)$, in Fig. \ref{fig:alpha_YIG}. }
\begin{tabular}{lccc}
\toprule
Curve name & Cutoff freq.~$f_{c2}$ & FMR freq.~$f_\mathrm{FMR}$ & Peak attenuation $\alpha_\mathrm{peak}$ \\
 & (GHz) & (GHz) & (dB) \\
\midrule
This work & 9.2717 & 10.4568 & 57.8862 \\
Experiment by \cite{gardiol1971} & 9.1605 & 10.4354 & 55.0005 \\
Theory 1 by \cite{gardiol1971} & 9.2388 & 10.2873 & 80.5410 \\
Theory 2 by \cite{uher1987} & 9.2485 & 10.3989 & 67.4374 \\
\bottomrule
\end{tabular}\\[10pt]
\label{tab:alpha_YIG}
\end{table*}

\begin{figure}[tb]
\centering
\includegraphics[width=0.51\textwidth]{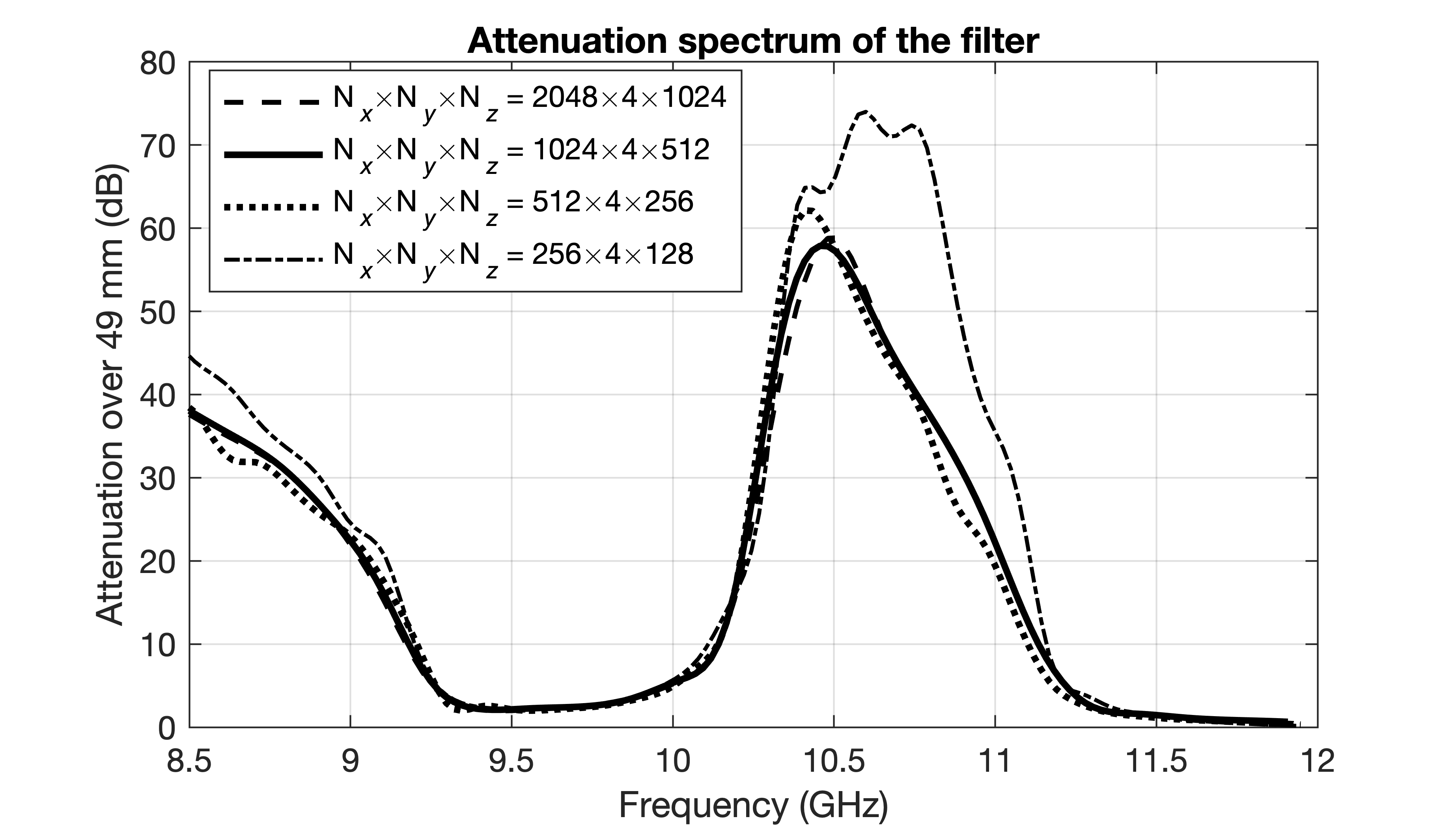}
\caption{Computational results for attenuation spectra of the magnetically tunable filter for varying resolutions.  We observe that by the 1024 $\times$ 4 $\times$ 512 case the solution does not significantly change with increasing resolution.}
\label{fig:mesh_converge}
\end{figure}

To verify grid-convergence, we performed simulations with different values of $\Delta x$ and $\Delta z$ (holding $\Delta y$ constant as the TE10-mode EM field is constant in $y$ direction), refining each by a factor of 2 with each simulation. 
The coarsest simulation has $256 \times 4 \times 128$ grid cells, and the finest simulation has $2048 \times 4 \times 1024$ cells.
All other numerical parameters remain the same, including the CFL coefficient being 0.8 for all cases.
The attenuation spectrum (analogous to Fig. \ref{fig:alpha_YIG}) for these four simulations are shown in Fig. \ref{fig:mesh_converge}.
We note that the coarsest simulation results for $\alpha(f)$ do not match the other results.  The curves for the two finest simulations exhibit little visible difference.
Therefore, we conclude that $1024\times 4\times 512$ grid cells used for comparison with theory and experiments is sufficient to capture the evolution of EM signals for this problem. 
%\MarginPar{Was dt the same for all these?  Or was the CFL the same?  I would also change the title in the figure from 'Meshing convergence' to something else.  Just say 'Attenuation spectrum of a magnetically tunable RF filter' again.}

\section{Conclusions and Future Work}\label{sec:conclusions}
We have developed an open-source, portable software for modeling a coupled LLG-Maxwell solver for microelectronic circuitry.
The algorithm is second-order in space and time with support for spatially varying material properties.
We demonstrated its scalability on the NERSC multicore and GPU partitions, including the GPU speedup compared to host CPU-only simulations on a node-by-node basis.
The agreement between experiments, theories and our simulations of realistic air-filled waveguide and magnetically tunable filter has demonstrated the code capability of predicting the coupled physics of EM and micromagnetics.

There are a number of algorithmic improvements we are planning to implement to increase the applicability of our solver to larger classes of problems in EM and micromagnetics.
These include adding spin exchange coupling, magnetic anisotropy, STT and spin-orbit-torque (SOT) in the LLG equation, to enable the software to model a wider range of spintronic devices such as MRAM, magnetic sensors, and magnetic RF front-ends proposed by \cite{suhl1956ire,Cui2019}.
%\MarginPar{Jackie please elaborate on the additional magnetization physics that can be added into our current model -- the torque terms.}
We are also interested in modeling quantum hardware that includes regions of superconducting material for capturing the non-linear behavior of Josephson Junctions.
We plan to explore various representations of superconductivity, ranging from the use of PEC conditions at the boundary to incorporating different constitutive models, such as the London equations.
To explore more advanced geometrical features typically found in circuits, in particular curved and non-axis-aligned interfaces, we will consider increasing levels of geometrical fidelity.
This can include a simple stair-step approach to define either one material type or another in each cell, to a blended volume weighted approach, to full inclusion of embedded boundaries developed by \cite{cary2006petascale}.
Embedded boundary, or ``cut-cell'' support is also available within the AMReX infrastructure, offering future possibilities for more advanced geometrical representation.
Embedded boundaries are being actively developed by the WarpX team, allowing us to leverage the discretizations and kernels in collaboration with the WarpX project.
We are also interested in incorporating spectral solvers such as PSATD to compare the possible tradeoffs between efficiency and accuracy.
Also, we can leverage the linear solvers provided by AMReX to develop or re-implement implicit approaches such as ADI.
AMReX also contains built-in geometric multigrid and Krylov-based linear solvers for elliptic and parabolic problems, as well as interfaces to external solvers available in the hypre and PETSc packages.
%\sout{Implicit discretizations can be developed to leverage these solvers for new classes of temporal integration strategies.}
AMReX also provides support for adaptive mesh refinement by managing communication between grids at different levels of refinement.
Incorporating AMR will offer increased spatial resolution in regions of interest.
The AMR support being developed for WarpX can be used as a springboard for our Maxwell+LLG solver, as well as future multiphysics coupling efforts.
We will consider the use of the AMReX particle class for other types of applications, such as phonon excitation proposed by \cite{Anufriev2017}.

\begin{acks}
This work was supported by the U.S.~Department of Energy, Office of Science, Office of Advanced Scientific Computing Research, Applied Mathematics Program under contract No.~DE-AC02-05CH11231.
This research used resources of the National Energy Research Scientific Computing Center, a DOE Office of Science User Facility supported by the Office of Science of the U.S. Department of Energy under Contract No.~DE-AC02-05CH11231.
This research leveraged the open-source particle-in-cell code WarpX \url{https://github.com/ECP-WarpX/WarpX} primarily funded by the US DOE Exascale Computing Project and the open source AMReX code, \url{https://github.com/AMReX-Codes/amrex}. We acknowledge all WarpX and AMReX contributors.
\end{acks}

\bibliographystyle{SageH}
\bibliography{LLG.bib}

\begin{thebibliography}{32}
\providecommand{\natexlab}[1]{#1}
\providecommand{\url}[1]{\texttt{#1}}
\providecommand{\urlprefix}{URL }
\expandafter\ifx\csname urlstyle\endcsname\relax
  \providecommand{\doi}[1]{DOI:\discretionary{}{}{}#1}\else
  \providecommand{\doi}{DOI:\discretionary{}{}{}\begingroup
  \urlstyle{rm}\Url}\fi

\bibitem[{Anufriev et~al.(2017)Anufriev, Ramiere, Maire and
  Nomura}]{Anufriev2017}
Anufriev R, Ramiere A, Maire J and Nomura M (2017) {Heat guiding and focusing
  using ballistic phonon transport in phononic nanostructures}.
\newblock \emph{Nature Communications} 8(1): 15505.
\newblock \doi{10.1038/ncomms15505}.
\newblock \urlprefix\url{https://doi.org/10.1038/ncomms15505}.

\bibitem[{Aziz(2009)}]{Aziz2009}
Aziz MM (2009) {Sub-nanosecond electromagnetic-micromagnetic dynamic
  simulations using the finite-difference time-domain method}.
\newblock \emph{Progress In Electromagnetics Research} 15: 1--29.

\bibitem[{Berenger(1996)}]{berenger1996three}
Berenger JP (1996) Three-dimensional perfectly matched layer for the absorption
  of electromagnetic waves.
\newblock \emph{Journal of computational physics} 127(2): 363--379.

\bibitem[{Berenger et~al.(1994)}]{berenger1994perfectly}
Berenger JP et~al. (1994) A perfectly matched layer for the absorption of
  electromagnetic waves.
\newblock \emph{Journal of computational physics} 114(2): 185--200.

\bibitem[{Cary et~al.(2006)Cary, Abell, Amundson, Bruhwiler, Busby, Carlsson,
  Dimitrov, Kashdan, Messmer, Nieter et~al.}]{cary2006petascale}
Cary JR, Abell D, Amundson J, Bruhwiler D, Busby R, Carlsson J, Dimitrov D,
  Kashdan E, Messmer P, Nieter C et~al. (2006) Petascale self-consistent
  electromagnetic computations using scalable and accurate algorithms for
  complex structures.
\newblock In: \emph{Journal of Physics: Conference Series}, volume~46. IOP
  Publishing, p. 027.

\bibitem[{Chappert et~al.(2007)Chappert, Fert and Van~Dau}]{chappert2007}
Chappert C, Fert A and Van~Dau F (2007) The emergence of spin electronics in
  data storage.
\newblock \emph{Nature Mater} 6: 813--823.

\bibitem[{Couture et~al.(2017)Couture, Chang, Volvach, Goncharov and
  Lomakin}]{couture2017coupled}
Couture S, Chang R, Volvach I, Goncharov A and Lomakin V (2017) Coupled
  finite-element micromagnetic—integral equation electromagnetic simulator
  for modeling magnetization—eddy currents dynamics.
\newblock \emph{IEEE Transactions on Magnetics} 53(12): 1--9.

\bibitem[{Cui et~al.(2019)Cui, Yao and Wang}]{Cui2019}
Cui H, Yao Z and Wang Y (2019) {Coupling Electromagnetic Waves to Spin Waves: A
  Physics-Based Nonlinear Circuit Model for Frequency-Selective Limiters}.
\newblock \emph{IEEE Transactions on Microwave Theory and Techniques} 67(8).
\newblock \doi{10.1109/TMTT.2019.2918517}.

\bibitem[{Donahue and Porter(1999)}]{OOMMF}
Donahue M and Porter D (1999) Oommf user's guide, version 1.0.
\newblock \emph{Interagency Report NISTIR 6376} .

\bibitem[{Fu et~al.(2015)Fu, Cui, Hu, Chang, Donahue and
  Lomakin}]{fu2015finite}
Fu S, Cui W, Hu M, Chang R, Donahue MJ and Lomakin V (2015) Finite-difference
  micromagnetic solvers with the object-oriented micromagnetic framework on
  graphics processing units.
\newblock \emph{IEEE Transactions on Magnetics} 52(4): 1--9.

\bibitem[{{Gardiol} and {Vander Vorst}(1971)}]{gardiol1971}
{Gardiol} FE and {Vander Vorst} AS (1971) Computer analysis of e-plane
  resonance isolators.
\newblock \emph{IEEE Transactions on Microwave Theory and Techniques} 19(3):
  315--322.
\newblock \doi{10.1109/TMTT.1971.1127505}.

\bibitem[{Garello et~al.(2013)Garello, Miron, Avci and et~al.}]{garello2013}
Garello K, Miron I, Avci C and et~al (2013) Symmetry and magnitude of
  spin–orbit torques in ferromagnetic heterostructures.
\newblock \emph{Nature Nanotech} 8: 587--593.

\bibitem[{Landau and Lifshitz(1992)}]{landau1992theory}
Landau L and Lifshitz E (1992) On the theory of the dispersion of magnetic
  permeability in ferromagnetic bodies.
\newblock In: \emph{Perspectives in Theoretical Physics}. Elsevier, pp. 51--65.

\bibitem[{Lax et~al.(1962)Lax, Button and Hagger}]{lax1962microwave}
Lax B, Button KJ and Hagger H (1962) \emph{Microwave ferrites and
  ferrimagnetics}.
\newblock MacGraw-Hill.

\bibitem[{Namiki(1999)}]{Namiki1999}
Namiki T (1999) {A New FDTD algorithm based on alternating-Direction implicit
  method}.
\newblock \emph{IEEE Transactions on Microwave Theory and Techniques}
  \doi{10.1109/22.795075}.

\bibitem[{Oti(1997)}]{oti1997simulmag}
Oti JO (1997) Simulmag version 1.0.
\newblock \emph{Electromagnetic Technology Division, NIST} .

\bibitem[{Pozar(2012)}]{Pozar2012}
Pozar DM (2012) \emph{{Microwave Engineering}}.
\newblock JohnWiley {\&} Sons, Inc.

\bibitem[{{Roy} and {Kailath}(1989)}]{esprit1}
{Roy} R and {Kailath} T (1989) Esprit-estimation of signal parameters via
  rotational invariance techniques.
\newblock \emph{IEEE Transactions on Acoustics, Speech, and Signal Processing}
  37(7): 984--995.
\newblock \doi{10.1109/29.32276}.

\bibitem[{Shapoval et~al.(2019)Shapoval, Vay and Vincenti}]{shapoval2019two}
Shapoval O, Vay JL and Vincenti H (2019) Two-step perfectly matched layer for
  arbitrary-order pseudo-spectral analytical time-domain methods.
\newblock \emph{Computer Physics Communications} 235: 102--110.

\bibitem[{Suhl(1956)}]{suhl1956ire}
Suhl H (1956) The nonlinear behavior of ferrites at high microwave signal
  levels.
\newblock \emph{Proceedings of IRE} 44(10): 1270--1284.

\bibitem[{Taflove and Hagness(2005)}]{taflove2005computational}
Taflove A and Hagness SC (2005) \emph{Computational electromagnetics: the
  finite-difference time-domain method}.
\newblock Artech House.

\bibitem[{{Uher} et~al.(1987){Uher}, {Arndt} and {Bornemann}}]{uher1987}
{Uher} J, {Arndt} F and {Bornemann} J (1987) Field theory design of
  ferrite-loaded waveguide nonreciprocal phase shifters with multisection
  ferrite or dielectric slab impedance transformers.
\newblock \emph{IEEE Transactions on Microwave Theory and Techniques} 35(6):
  552--560.
\newblock \doi{10.1109/TMTT.1987.1133703}.

\bibitem[{Vacus and Vukadinovic(2005)}]{Vacus2005}
Vacus O and Vukadinovic N (2005) {Dynamic susceptibility computations for thin
  magnetic films}.
\newblock \emph{Journal of Computational and Applied Mathematics} 176(2):
  263--281.
\newblock \doi{http://dx.doi.org/10.1016/j.cam.2004.07.016}.
\newblock \urlprefix\url{internal-pdf://140.18.172.127/Dynamic Susceptibility
  computations for thin m.pdf
  http://www.sciencedirect.com/science/article/pii/S0377042704003322}.

\bibitem[{Vansteenkiste et~al.(2014)Vansteenkiste, Leliaert, Dvornik, Helsen,
  Garcia-Sanchez and Van~Waeyenberge}]{mumax}
Vansteenkiste A, Leliaert J, Dvornik M, Helsen M, Garcia-Sanchez F and
  Van~Waeyenberge B (2014) The design and verification of mumax3.
\newblock \emph{AIP advances} 4(10): 107133.

\bibitem[{Vay et~al.(2018)Vay, Almgren, Bell, Ge, Grote, Hogan, Kononenko,
  Lehe, Myers, Ng et~al.}]{vay2018warp}
Vay JL, Almgren A, Bell J, Ge L, Grote D, Hogan M, Kononenko O, Lehe R, Myers
  A, Ng C et~al. (2018) {Warp-X: A new exascale computing platform for
  beam--plasma simulations}.
\newblock \emph{Nuclear Instruments and Methods in Physics Research Section A:
  Accelerators, Spectrometers, Detectors and Associated Equipment} 909:
  476--479.

\bibitem[{Venugopal et~al.(2019)Venugopal, Qu and
  Victora}]{venugopal2019nonlinear}
Venugopal A, Qu T and Victora R (2019) Nonlinear parallel-pumped fmr: Three and
  four magnon processes.
\newblock \emph{IEEE Transactions on Microwave Theory and Techniques} 68(2):
  602--610.

\bibitem[{{Wang} and {Ling}(1998)}]{esprit2}
{Wang} Y and {Ling} H (1998) Multimode parameter extraction for multiconductor
  transmission lines via single-pass fdtd and signal-processing techniques.
\newblock \emph{IEEE Transactions on Microwave Theory and Techniques} 46(1):
  89--96.
\newblock \doi{10.1109/22.654927}.

\bibitem[{Yao et~al.(2018)Yao, Tok, Itoh and Wang}]{yao2018multiscale}
Yao Z, Tok R, Itoh T and Wang Y (2018) {A multiscale unconditionally stable
  time-domain (MUST) solver unifying electrodynamics and micromagnetics}.
\newblock \emph{IEEE Transactions on Microwave Theory and Techniques} 66(6):
  2683--2696.

\bibitem[{Yee(1966)}]{yee1966numerical}
Yee K (1966) Numerical solution of initial boundary value problems involving
  maxwell's equations in isotropic media.
\newblock \emph{IEEE Transactions on antennas and propagation} 14(3): 302--307.

\bibitem[{Zhang et~al.(2019)Zhang, Almgren, Beckner, Bell, Blaschke, Chan, Day,
  Friesen, Gott, Graves et~al.}]{zhang2019amrex}
Zhang W, Almgren A, Beckner V, Bell J, Blaschke J, Chan C, Day M, Friesen B,
  Gott K, Graves D et~al. (2019) {AMReX}: {A} framework for block-structured
  adaptive mesh refinement.
\newblock \emph{Journal of Open Source Software} 4(37): 1370--1370.

\bibitem[{Zhang et~al.(2020)Zhang, Almgren, Beckner, Bell, Blaschke, Chan, Day,
  Friesen, Gott, Graves et~al.}]{zhang2020amrex}
Zhang W, Almgren A, Beckner V, Bell J, Blaschke J, Chan C, Day M, Friesen B,
  Gott K, Graves D et~al. (2020) {AMReX}: {B}lock-structured adaptive mesh
  refinement for multiphysics applications.
\newblock \emph{submitted for publication} .

\bibitem[{Zheng and Chen(2000)}]{Zheng2000}
Zheng F and Chen Z (2000) {Toward the development of a three-dimensional
  unconditionally stable finite-difference time-domain method}.
\newblock \emph{IEEE Transactions on Microwave Theory and Techniques}
  \doi{10.1109/22.869007}.

\end{thebibliography}

\end{document}